\documentclass[11pt,a4paper]{article}
\usepackage{amsfonts}
\usepackage{amsmath}
\usepackage{amssymb}
\usepackage{amsthm}

\usepackage{geometry}
\usepackage{indentfirst}
\usepackage{graphicx} % Required for inserting images
\usepackage{natbib} \bibpunct{(}{)}{;}{a}{,}{,}
\usepackage[colorlinks=true,urlcolor=blue,citecolor=blue,linkcolor=blue,bookmarks=true]{hyperref}
\usepackage{appendix}
\usepackage{mathrsfs}
\usepackage{enumerate}
\usepackage{tikz}

\usepackage{subcaption}
\usepackage[labelfont=bf]{caption}
\usepackage{booktabs,color,url}
\usepackage{diagbox}

\theoremstyle{plain}
\newtheorem{theorem}{Theorem}[section]

\newtheorem{proposition}{Proposition}[section]
\newtheorem{lemma}{Lemma}[section]
\newtheorem{corollary}{Corollary}[section]
\newtheorem{assumption}{Assumption}[section]
\newtheorem{definition}{Definition}[section]
\newtheorem{remark}{Remark}[section]

\DeclareMathOperator*{\argmin}{arg\,min}
\allowdisplaybreaks

\renewcommand{\qed}{%
  \unskip\nobreak\hfill\penalty50
  \hskip1em\null\nobreak\hfill
  \mbox{\rule{6pt}{6pt}}\par
}

\renewenvironment{proof}
  {\par\noindent\textbf{Proof:}\enspace\ignorespaces}
  {\qed}

\theoremstyle{definition}

\newcommand{\R}{\mathbb{R}}

\newcommand{\Rmnum}[1]{\uppercase\expandafter{\romannumeral #1\relax}}

\newcommand{\VaR}{\mathrm{VaR}}
\newcommand{\ES}{\mathrm{ES}}
\newcommand{\RVaR}{\mathrm{RVaR}}
\def\d{\mathrm{d}}
\title{Robust distortion riskmetrics under symmetry, unimodality, and Wasserstein constraints}
\date{\today}
\author{Peng Liu\thanks{Department of Mathematics, Statistics and Actuarial Science, University of Essex, Colchester, UK. Email: \url{peng.liu@essex.ac.uk}}\and Yiying Zhang\thanks{Department of Mathematics, Southern University of Science and Technology, Shenzhen, P.R. China. Email: \url{zhangyy3@sustech.edu.cn}} \and Baishuai Zuo\thanks{Department of Mathematics, Southern University of Science and Technology, Shenzhen,  P.R. China. Email: \url{bszuo124@163.com}}}

\newif\ifshowrevisions
\showrevisionsfalse
\newcommand{\revisioncolor}{%
  \ifshowrevisions
    \color{red}%
    \hypersetup{linkcolor=red,citecolor=red,urlcolor=red}%
  \fi}

\definecolor{studybrown}{rgb}{0.6,0.4,0.2}

\newcommand{\studyedit}[1]{\begingroup\ifshowrevisions\color{studybrown}\fi#1\endgroup}
\DeclareRobustCommand{\notationedit}[1]{\begingroup\ifshowrevisions\color{red}\fi#1\endgroup}
\newcommand{\simGridFeasiblePairs}{131}

\newcommand{\simNormalBudgetSmall}{0.0315}

\newcommand{\simNormalBudgetLarge}{0.0042}

\newcommand{\simNormalSymmetricReduction}{6.7}
\newcommand{\simNormalUnimodalReduction}{0.6}

\newcommand{\simStudentSymmetricCoverageLarge}{96.7}
\newcommand{\simStudentJointCoverageLarge}{95.8}
\newcommand{\simStudentSymmetricReduction}{10.0}
\newcommand{\simStudentUnimodalReduction}{0.9}

\newcommand{\simBimodalSymmetricFeasible}{100.0}

\newcommand{\simBimodalJointFeasible}{1.6}

\newcommand{\simLognormalCalibrationSuccess}{78.7}

\newcommand{\simLognormalGeneralCoverage}{78.7}

\newcommand{\simGridChange}{5.58\times10^{-4}}

\begin{document}

\maketitle

\begin{abstract}
We study worst-case distortion riskmetrics over symmetric distributions with prescribed mean and variance in a Wasserstein ball around a reference distribution. Our key contribution is the construction of a reference-dependent auxiliary distortion that encodes the Wasserstein constraint while preserving symmetry under the concave-envelope operation. Under suitable regularity and nondegeneracy conditions, this construction yields sharp upper bounds for distortions of bounded variation, allowing nonmonotonicity and interior discontinuities, and identifies extremal distributions when the bounds are attained. An approximation preserving symmetry and the prescribed moments exactly shows that upper semicontinuous relaxation of the symmetrized distortion leaves the supremum unchanged for budgets strictly above the feasibility threshold. With additional unimodality, symmetry fixes the quantile's concavity–convexity transition at the midpoint, yielding a characterization via $L^2$-projection onto a fixed closed convex cone for absolutely continuous distortions with square-integrable derivatives. Numerical examples for ES, RVaR, and VaR illustrate how shape restrictions and the reference distribution affect feasibility and worst-case risk. Simulations with data-dependent Wasserstein radii examine worst-case ES and feasibility under correctly specified and misspecified shape restrictions.
\\[1mm]

\noindent


\end{abstract}
\section{Introduction}\label{sec:intro}
Risk assessment typically relies on a probability distribution estimated
from data or specified through modeling assumptions. The resulting risk
estimates can therefore be sensitive to estimation error and model
misspecification, concerns recognized in both regulatory and actuarial
discussions; see \cite{BASEL19,Actuarial2017comments}. A standard way to
account for this uncertainty is to evaluate the worst-case value of a
risk functional over a set of plausible distributions. For a risk
functional $\rho$, this leads to
\begin{align}\label{x2}
    \sup_{F\in\mathcal{M}}\rho(F),
\end{align}
where $\mathcal{M}$ consists of distributions on $\mathbb R$ that are
consistent with the available information. This maximization also
arises as the inner problem in distributionally robust optimization.

This paper focuses on distortion riskmetrics, a class that includes
both distortion risk measures and measures of variability. Examples
include Value-at-Risk (VaR), Expected Shortfall (ES), Range Value-at-Risk
(RVaR), Gini deviation, and mean-median deviation. These functionals
are used in regulatory risk assessment \citep{BASEL19} and insurance
pricing \citep{Denuit2005actuarial}, and are related to statistical
$L$-functionals \citep{Huber2009robust}, dual utility theory
\citep{Yaari1987dual}, and spectral risk measures
\citep{Acerbi2002spectral}. Distortion riskmetrics allow nonmonotone
and discontinuous distortion functions; see
\cite{Wang2020distortion,Wang2020characterization}. For a general
treatment of risk measures, we refer to
\citet[Chapter~4]{follmer2016stochastic}.

Prescribing the mean and variance leaves considerable freedom in the shape of a distribution. Symmetry and unimodality restrict this freedom without imposing a parametric family, while a Wasserstein ball limits departures from a reference distribution. We study these restrictions jointly using the $2$-Wasserstein distance, focusing on how the worst-case risk varies with the distance budget when the moment and shape constraints are held fixed.

Symmetry is a natural assumption when deviations on either side of a central value are approximately balanced, as in some measurement-error models \citep{fuller1987measurement}. Familiar examples include the normal, Laplace, logistic, and uniform distributions, as well as Student's $t$ distributions with finite variance. Even when the reference distribution is symmetric, however, the distributions in a Wasserstein ball need not be. Imposing symmetry explicitly may therefore sharpen the worst-case bound. We also examine the effect of imposing unimodality in addition to symmetry.

Under prescribed first and second moments, \citet{ElGhaoui2003worst} derived closed-form bounds for VaR. Related results are available for ES \citep{Chen2011tight} and RVaR \citep{li2018worst}, while analytical bounds for spectral risk measures were established by \citet{Li2018technical}. For a general class of distortion risk measures, \citet{Cai2025distributionally} studied worst-case distributions and their interpretation in terms of risk aversion. The analysis in \citet{Pesenti2025robust} covers nonconvex distortion riskmetrics over general uncertainty sets, including those defined by moment information.

Shape restrictions have also been used to refine moment-based bounds. For RVaR, \citet{li2018worst} considered symmetry, unimodality, and their combination, and \citet{Bernard2020range} obtained upper and lower bounds under moment and unimodality constraints. Symmetry and unimodality were also considered by \citet{bernard2025model}. For more general distortion risk measures, \citet{shao2023distortion,shao2024extreme} derived extremal bounds under symmetry and moment constraints, including higher-order moment information. \citet{Zuo2025analzing} established sharp bounds for distortion riskmetrics under moment and symmetry constraints, as well as upper and lower bounds under unimodality, with or without symmetry. We impose a Wasserstein constraint alongside these moment and shape restrictions. The reference distribution then affects both the minimum budget required for feasibility and the threshold beyond which enlarging the ball no longer changes the worst-case value.

For uncertainty sets defined by Wasserstein balls, \citet{LMWW22} derived sharp upper bounds for distortion risk measures with absolutely continuous distortion functions, while \citet{liu2026robust} treated general distortion riskmetrics. The combination of a Wasserstein ball and prescribed mean and variance was studied by \citet{bernard2024robust}, who used isotonic projection to characterize extremal values for increasing, absolutely continuous distortion functions. Under similar constraints, \citet{Cai2024worst} considered distortion risk measures of transformed risks.

Our analysis builds most directly on \citet{liu2025robust}, who derived sharp upper and lower bounds for distortion riskmetrics under mean, variance, and Wasserstein constraints. Their results apply to distortion functions of bounded variation, allowing for discontinuities and nonmonotonicity. Their approach relies on the concave envelope of a linear combination of the distortion function and a functional of the reference distribution. For absolutely continuous distortions, they also incorporated unimodality, developed computational procedures, and studied robust portfolio optimization.

\begin{table}[htbp]
\centering
\footnotesize
\caption{Distributional information considered in selected studies on
worst-case risk assessment. A checkmark indicates that the information
is incorporated in at least one formulation in the cited study; a cross
indicates that it is not considered.}
\label{tab:lit}
\begin{tabular}{llcccc}
\toprule
Reference & Risk functional & First two moments & Wasserstein & Symmetry & Unimodality\\
\midrule
\cite{ElGhaoui2003worst} & VaR
  & $\checkmark$ & $\times$ & $\times$ & $\times$ \\
\cite{Chen2011tight} & TVaR
  & $\checkmark$ & $\times$ & $\times$ & $\times$ \\
\cite{li2018worst} & RVaR
  & $\checkmark$ & $\times$ & $\checkmark$ & $\checkmark$ \\
\cite{Li2018technical} & Spectral risk measures
  & $\checkmark$ & $\times$ & $\times$ & $\times$ \\
\cite{Bernard2020range} & RVaR
  & $\checkmark$ & $\times$ & $\times$ & $\checkmark$ \\
\cite{LMWW22} & Distortion risk measures
  & $\times$ & $\checkmark$ & $\times$ & $\times$ \\
\cite{shao2023distortion} & Distortion risk measures
  & $\checkmark$ & $\times$ & $\checkmark$ & $\times$ \\
\cite{bernard2024robust} & Distortion risk measures
  & $\checkmark$ & $\checkmark$ & $\times$ & $\times$ \\
\cite{Cai2024worst} & Stop-loss, limited-loss
  & $\checkmark$ & $\checkmark$ & $\times$ & $\times$ \\
\cite{shao2024extreme} & Distortion risk measures
  & $\checkmark$ & $\times$ & $\checkmark$ & $\times$ \\
\cite{bernard2025model} & RVaR
  & $\checkmark$ & $\times$ & $\checkmark$ & $\checkmark$ \\
\cite{Cai2025distributionally} & Distortion risk measures
  & $\checkmark$ & $\times$ & $\times$ & $\times$ \\
\cite{liu2025robust} & Distortion riskmetrics
  & $\checkmark$ & $\checkmark$ & $\times$ & $\checkmark$ \\
\cite{Pesenti2025robust} & Distortion riskmetrics
  & $\checkmark$ & $\times$ & $\times$ & $\times$ \\
\cite{liu2026robust} & Distortion riskmetrics
  & $\times$ & $\checkmark$ & $\times$ & $\times$ \\
This paper & Distortion riskmetrics
  & $\checkmark$ & $\checkmark$ & $\checkmark$ & $\checkmark$ \\
\bottomrule
\end{tabular}
\end{table}

In this paper, we study worst-case distortion riskmetrics under
joint moment, symmetry, and Wasserstein constraints, both with
and without an additional unimodality restriction. The interaction
between symmetry and Wasserstein ambiguity distinguishes our
analysis from the existing frameworks reviewed above, particularly
in the characterization of feasibility and extremal distributions.
Table~\ref{tab:lit} compares the distributional information
incorporated in selected related studies and highlights these
differences. The contributions of our paper is as follows:

First, we construct a reference-dependent auxiliary distortion
that accommodates the moment and Wasserstein constraints while
preserving symmetry under the concave-envelope operation.
Under the stated regularity and nondegeneracy assumptions,
this construction yields sharp upper bounds and characterizes
extremal distributions when the bounds are attained.
We also characterize the unique nearest symmetric distribution
with the prescribed moments through the antisymmetric part of
the reference quantile. Its distance from the reference determines
the feasibility threshold, and an exact distance decomposition
recasts the Wasserstein constraint around this nearest
distribution with a rescaled budget.

Second, for discontinuous distortions, we use approximations
that preserve symmetry and the first two moments exactly to
show that upper semicontinuous relaxation of the symmetrized
distortion leaves the supremum unchanged for budgets strictly
above the feasibility threshold. Under our assumptions on the
symmetric uncertainty class, the proof requires no separate
continuity condition on the relaxed risk along the multiplier
path, such as that imposed in
\citet[Theorem~2(i)]{liu2025robust}. We also derive explicit
formulas for ES, RVaR, VaR, Gini deviation, and mean-median
deviation. For left VaR, we prove nonattainment throughout
the active-budget region and, at upper-tail probability
levels, for every budget strictly above the feasibility
threshold.

Third, under joint symmetry and unimodality, the quantile's
concavity-convexity transition is fixed at $1/2$. For
absolutely continuous distortions with square-integrable
derivatives, this yields an $L^2$-projection characterization
on a fixed closed convex cone and a one-dimensional dual
representation. We characterize the worst-case quantile
and value and obtain an explicit solution for Gini's premium
principle under a symmetric unimodal reference. For computation,
we develop approximations based on convex quadratic programs
on fixed grids, with error bounds for the projected quantiles
and risk values.

Numerical examples for ES, RVaR, and VaR illustrate how shape
restrictions affect feasibility and worst-case risk, and how
the bound becomes independent of the reference distribution
beyond the saturation threshold. \studyedit{Simulations
compare worst-case ES and feasibility at common calibrated
Wasserstein budgets under correctly specified and misspecified
shape restrictions, distinguishing the bootstrap validation
target from population ES coverage.}

The remainder of the paper is outlined as follows. Section~\ref{sec2} introduces the definitions and optimization problems.
Section~\ref{sec3} first develops symmetry reduction and feasibility,
then states the general sharp-bound and relaxation theorems, and finally
specializes them to explicit riskmetrics. Section~\ref{sec4} develops
the projection method under joint symmetry and unimodality.
Section~\ref{sec5} presents numerical examples \studyedit{and a
simulation study of radius calibration and worst-case ES.
Section~\ref{sec:conclusion}} concludes. The proofs for Sections~\ref{sec3} and~\ref{sec4} are collected
in Appendices~\ref{app:symmetry} and~\ref{app:unimodality}, respectively.

\section{Problem formulation}\label{sec2}
We work on an atomless probability space \((\Omega,\mathcal{F},\mathbb{P})\). Let \(\mathcal{M}^2\) denote the set of distribution functions of real-valued random variables with finite second moments. For \(G\in\mathcal{M}^2\), define its left and right quantile functions by
\[
\begin{aligned}
G^{-1}(p)
&:=\inf\{x\in\mathbb{R}:G(x)\ge p\},
&&p\in(0,1],\\
G^{-1+}(p)
&:=\sup\{x\in\mathbb{R}:G(x)\le p\},
&&p\in[0,1),
\end{aligned}
\]
respectively, with the endpoint conventions \(G^{-1}(0):=G^{-1+}(0)\) and \(G^{-1+}(1):=G^{-1}(1)\). Endpoint values are understood in the extended real sense.
We use the left quantile to define VaR and write
 $\operatorname{VaR}_p(G):=G^{-1}(p)$, $\operatorname{VaR}_p^{+}(G):=G^{-1+}(p)$, for $p\in[0,1]$. Throughout, \(U\) denotes a random variable uniformly distributed on \((0,1)\).
\notationedit{For a real number $x$, $|x|$ denotes its absolute value.}

\subsection{Distortion riskmetrics}\label{subsec2.1}
Let
\[
\mathcal{G}
=
\left\{
g:[0,1]\to\mathbb{R}:
g(0)=0,\;
g \text{ is of bounded variation}
\right\}.
\]
For $g\in\mathcal{G}$, the associated distortion riskmetric is defined by
\begin{align}\label{b2}
\rho_g(G)
=
\int_{-\infty}^{0}
\left[g\bigl(1-G(x)\bigr)-g(1)\right]\mathrm{d}x
+
\int_{0}^{\infty}
g\bigl(1-G(x)\bigr)\mathrm{d}x,
\end{align}
whenever both integrals are finite; see \cite{Wang2020distortion}.
The function $g$ is called the distortion function and need not be
monotone or continuous. If $g$ is absolutely continuous and
$g'(1-\cdot)G^{-1}(\cdot)\in L^1(0,1)$, then
\begin{align}\label{(a)}
\rho_g(G)
=
\int_{0}^{1}g'(1-u)G^{-1}(u)\mathrm{d}u,
\end{align}
where $g'$ denotes the almost-everywhere derivative of $g$.

The following examples will be used throughout the paper.

\begin{enumerate}[(i)]

\item \textit{Range Value-at-Risk (\text{RVaR}).}
For $0<\alpha<\beta<1$,
\[
\mathrm{RVaR}_{\alpha,\beta}(G)
=
\frac{1}{\beta-\alpha}
\int_{\alpha}^{\beta}G^{-1}(u)\mathrm{d}u,
\]
with distortion
$g(t)=\min\{1,\max\{0,(t+\beta-1)/(\beta-\alpha)\}\}$.
VaR and ES arise as the limits
$\lim_{\beta\uparrow\alpha}
\mathrm{RVaR}_{\beta,\alpha}(G)=\mathrm{VaR}_{\alpha}(G)$
and
$\lim_{\beta\uparrow1}
\mathrm{RVaR}_{\alpha,\beta}(G)=\mathrm{ES}_{\alpha}(G)$.
For background and applications, see
\cite{cont2010robustness,embrechts2018quantile,blanchet2025convolution}.

\item \textit{Gini deviation (GD) and Gini's premium principle.}
For independent random variables $X,Y\sim G$, we use the normalization
\[
\mathrm{GD}(G)
:=
\frac{1}{2}\mathbb{E}[|X-Y|]
=
\rho_g(G),
\qquad
g(t)=t(1-t).
\]
Gini's distortion premium principle is then given by
\[
\mathrm{GDP}_r(G)
:=
\mathbb{E}[X]+r\,\mathrm{GD}(G)
=
\rho_{g_r}(G),
\qquad
g_r(t)=(1+r)t-rt^2,
\quad 0\le r\le1.
\]
These functionals are used in portfolio optimization, risk sharing,
and insurance pricing; see
\cite{shalit1984mean,Lauzier2025risk,Denneberg1990premium,Wang1996premium}.

\item \textit{Mean-median deviation (MMD).}
For $X\sim G$,
\[
\mathrm{MMD}(G)
:=
\min_{a\in\mathbb{R}}\mathbb{E}[|X-a|]
=
\rho_g(G),
\qquad
g(t)=\min\{t,1-t\}.
\]
This measure quantifies dispersion around a median; see
\cite{Denneberg1990premium,Lauzier2025risk}.

\end{enumerate}

\subsection{Modeling distributional uncertainty}\label{subsec2.2}

We now define the shape restrictions and distance used in the two
uncertainty sets. For unimodality, we adopt the classical definition
\notationedit{\citep{Khintchine1938unimodal,Olshen1970generalized,dharmadhikari1988unimodality}}.

\begin{definition}[Unimodality]\label{def:unimodality}
A distribution function $G\in\mathcal{M}^2$ is unimodal if there exists
a mode $m\in\mathbb{R}$ such that $G$ is convex on $(-\infty,m)$ and
concave on $(m,\infty)$. An atom at $m$ is permitted.
\end{definition}

We use the equivalent quantile characterization in
\citet[Lemma~2]{liu2025robust}.

\begin{lemma}[Quantile characterization of unimodality]\label{le.a1}
A distribution function $G$ is unimodal if and only if $G^{-1}$ is
continuous on $(0,1)$ and there exists $\varsigma\in[0,1]$ such that
$G^{-1}$ is concave on $(0,\varsigma)$ and convex on $(\varsigma,1)$.
The cases $\varsigma=0$ and $\varsigma=1$ correspond to convexity and
concavity on the whole interval, respectively.
\end{lemma}

If $G$ is also symmetric, the transition point in Lemma~\ref{le.a1}
can be chosen as $\varsigma=1/2$. Thus its quantile is concave on
$(0,1/2)$ and convex on $(1/2,1)$.
The fixed transition at $1/2$ is the basis of the projection
method in Section~\ref{sec4}.

Proximity to the reference distribution is measured by the
$2$-Wasserstein distance.

\begin{definition}[$2$-Wasserstein distance]\label{def:wasserstein}
For $G_1,G_2\in\mathcal{M}^2$, the $2$-Wasserstein distance is
\begin{equation}\label{eq:wasserstein-distance}
d_W(G_1,G_2)
=
\left(
\int_0^1
\left|G_1^{-1}(u)-G_2^{-1}(u)\right|^2\,\mathrm{d}u
\right)^{1/2}.
\end{equation}
\end{definition}

See \cite{Villani2009optimal} for the optimal transport background,
and \cite{MohajerinEsfahaniKuhn2018,BM19} for applications to distributionally robust optimization.

Let $\mathcal{M}^s$ denote the class of symmetric distributions in
$\mathcal{M}^2$, and set
\[
\mathcal{M}^{us}
:=
\{G\in\mathcal{M}^s:G\text{ is unimodal}\}.
\]
{
To add moment and distance constraints, fix a reference distribution
}
$F\in\mathcal{M}^2$, a mean $\mu\in\mathbb{R}$, a standard deviation
$\sigma>0$, and a Wasserstein budget $\epsilon\geq0$. The first uncertainty
set is
\begin{equation}\label{eq:symmetric-ambiguity}
\mathcal{M}_{\epsilon}^{s}(\mu,\sigma)
:=
\left\{
G\in\mathcal{M}^s:
\begin{array}{l}
\displaystyle \int_{\mathbb{R}}x\,\mathrm{d}G(x)=\mu,\qquad
\displaystyle \int_{\mathbb{R}}(x-\mu)^2\,\mathrm{d}G(x)=\sigma^2,\\[1ex]
d_W^2(G,F)\le\epsilon
\end{array}
\right\}.
\end{equation}
Here $d_W$ is defined in \eqref{eq:wasserstein-distance}, so the
Wasserstein radius is $\sqrt{\epsilon}$.
The reference distribution $F$ is arbitrary in $\mathcal{M}^2$.
For a distribution with mean $\mu$, symmetry is equivalently expressed as
\begin{equation}\label{eq:quantile-symmetry}
G^{-1}(u)+G^{-1+}(1-u)=2\mu
\qquad{\text{for every }u\in(0,1).}
\end{equation}
When $G$ is also unimodal, Lemma~\ref{le.a1} ensures that the two
quantile versions coincide on $(0,1)$.

The second uncertainty set adds unimodality to
\eqref{eq:symmetric-ambiguity}:
\begin{equation}\label{eq:unimodal-ambiguity}
\mathcal{M}_{\epsilon}^{us}(\mu,\sigma)
:=
\mathcal{M}_{\epsilon}^{s}(\mu,\sigma)\cap\mathcal{M}^{us}.
\end{equation}
For $j\in\{s,us\}$, removing the Wasserstein constraint gives
\[
\mathcal{M}^{j}(\mu,\sigma)
:=
\mathcal{M}_{\infty}^{j}(\mu,\sigma)
=
\left\{
G\in\mathcal{M}^{j}:
\int_{\mathbb{R}}x\,\mathrm{d}G(x)=\mu,\quad
\int_{\mathbb{R}}(x-\mu)^2\,\mathrm{d}G(x)=\sigma^2
\right\}.
\]
For any class $\mathcal{A}$ of distributions, write
$\mathcal{A}^{-1}:=\{G^{-1}:G\in\mathcal{A}\}$.
In particular, we use
$\mathcal{M}^{j,-1}$,
$\mathcal{M}^{j,-1}(\mu,\sigma)$, and
$\mathcal{M}_{\epsilon}^{j,-1}(\mu,\sigma)$
for the corresponding quantile classes.
In integrals and projections, we regard quantiles as elements of
$L^2(0,1)$ and identify functions that agree almost everywhere.
For pointwise formulas and VaR evaluations, we use the left-continuous
version. The notation $G^{-1+}$ always denotes the right-continuous version.
The class $\mathcal{M}^{us,-1}$, with no moment restrictions,
is the cone used for the projections in Section~\ref{sec4}.

For a nonempty uncertainty set $\mathcal M$ and $g\in\mathcal G$, we study
\begin{align}\label{pm1}
\sup_{G\in\mathcal{M}}\rho_g(G),
\end{align}
where $\rho_g$ is defined in \eqref{b2} and
$\mathcal{M}$ is either \eqref{eq:symmetric-ambiguity} or
\eqref{eq:unimodal-ambiguity}.
The assumptions needed for finiteness and attainment are stated
with each result. The two uncertainty sets are treated in
Sections~\ref{sec3} and~\ref{sec4}, respectively.

Since $-g\in\mathcal{G}$ and $\rho_{-g}=-\rho_g$, the corresponding
best-case problem satisfies
\begin{align}\label{f1}
\inf_{G\in\mathcal{M}}\rho_g(G)
=
-\sup_{G\in\mathcal{M}}\rho_{-g}(G).
\end{align}
A best-case bound therefore follows by applying the corresponding
worst-case result to $-g$, provided its assumptions hold for $-g$.

Section~\ref{sec3} uses the concave envelope, or least concave
majorant, of $g\in\mathcal G$:
\begin{equation}\label{eq:concave-envelope}
g^*(t)
:=
\inf\left\{
h(t):
h:[0,1]\to\mathbb{R}\text{ is concave and }h\ge g
\right\},
\qquad t\in[0,1].
\end{equation}
We also use the greatest convex minorant of $g$, defined by
\begin{equation}\label{eq:convex-envelope}
g_*(t)
:=
\sup\left\{
h(t):
h:[0,1]\to\mathbb R\text{ is convex and }h\leq g
\right\},
\qquad t\in[0,1].
\end{equation}
For the analysis in Section~\ref{sec3}, define the regularity class
\begin{equation}\label{eq:regular-distortions}
\mathcal G_2
:=
\left\{
g\in\mathcal G:
\begin{array}{l}
g(0+)=g(0)=0,\quad g(1-)=g(1),\\[2pt]
(g^*)',\ (g_*)'\in L^2(0,1)
\end{array}
\right\}.
\end{equation}
Endpoint continuity is inherited by the concave and convex envelopes;
see \citet[Section~2.2]{Pesenti2025robust}. A finite concave or convex
function continuous on $[0,1]$ is absolutely continuous. Thus both
envelopes of $g\in\mathcal G_2$ admit integral representations in terms
of their derivatives. Proposition~6 of \cite{Pesenti2025robust},
with $p=q=2$, then ensures that $\rho_g(G)$ is finite for every
$G\in\mathcal M^2$.
Functions in $\mathcal G_2$ may be nonmonotone and discontinuous
in the interior. The identities $(-g)^*=-g_*$ and $(-g)_*=-g^*$
show that $-g\in\mathcal G_2$ whenever $g\in\mathcal G_2$.
Thus finiteness is preserved in the best-case reduction \eqref{f1}.
Throughout, $\mathbf{1}_A$ denotes the indicator of a set $A$, and
$a\vee b:=\max\{a,b\}$ and $a\wedge b:=\min\{a,b\}$.
For a concave or convex function $g$, we use
$g'(t):=\partial_+g(t)$ for its right derivative at $t\in(0,1)$.
\notationedit{We write $\partial_-g(t)$ for its left derivative.}
Thus $t\mapsto(g_\lambda^*)'(1-t)$ in the quantile constructions below
is left-continuous. In integrals, either one-sided derivative gives
the same value.
The left- and right-hand limits at an interior point $x$ are denoted
by $g(x-)$ and $g(x+)$, respectively.

\section{Robust distortion riskmetrics under symmetry}\label{sec3}

\subsection{Symmetrization and feasibility}\label{subsec:sym-feasibility}

\notationedit{Let $\mu_F$ and $\sigma_F$ denote the mean and standard
deviation of the reference distribution $F$. For $f,h\in L^2(0,1)$,
$\langle f,h\rangle:=\int_0^1 f(u)h(u)\,du$ and
$\|f\|:=\langle f,f\rangle^{1/2}$ denote the inner product and norm.
For a square-integrable random variable $X$, we similarly write
$\|X\|:=(\mathbb E[|X|^2])^{1/2}$.}
Assume throughout this section that the reference distribution is
nondegenerate, with $\sigma_F>0$; we discuss the degenerate case below.
For $g\in\mathcal G$, define the symmetrized distortion
$\bar{g}(t)=\frac{g(t)+g(1-t)-g(1)}{2}$, so that
$\bar{g}(0)=\bar{g}(1)=0$.
The following identity expresses the risk of a symmetric distribution
in terms of $\bar g$.

\begin{proposition}\label{prop.1} For $g\in\mathcal G$ and a distribution $G$ symmetric about its mean $\mu$, if $\rho_{g}(G)$ is finite, then
\begin{align*}
\rho_{g}(G)=\mu g(1)+\rho_{\bar{g}}(G).
\end{align*}
\end{proposition}

We first determine when the uncertainty set is nonempty. Write
\[
s_F(u)=\frac{F^{-1}(u)-{F^{-1+}(1-u)}}2,\qquad
\sigma_0=\notationedit{\|s_F\|}>0,\qquad
A_F=(\mu_F-\mu)^2+\sigma_F^2+\sigma^2.
\]
\begin{lemma}[Feasibility under symmetry]\label{lem:sym-feasibility}
Set $\epsilon_0=A_F-2\sigma\sigma_0$, and let $H_\infty$ have quantile
\begin{equation}\label{eq:nearest-symmetric-quantile}
H_\infty^{-1}(u)=\mu+\frac{\sigma}{\sigma_0}s_F(u),\qquad u\in(0,1).
\end{equation}
Then $\mathcal M_\epsilon^s(\mu,\sigma)$ is empty for
$\epsilon<\epsilon_0$, equals $\{H_\infty\}$ for
$\epsilon=\epsilon_0$, and contains more than one distribution for
$\epsilon>\epsilon_0$. At the minimum budget, the worst-case value
is therefore $\rho_g(H_\infty)$ whenever $g\in\mathcal G_2$.
\end{lemma}
For every $G\in\mathcal M^s(\mu,\sigma)$, the nearest distribution
$H_\infty$ also gives the exact decomposition
\begin{equation}\label{eq:symmetric-budget-geometry}
d_W^2(F,G)
=\epsilon_0+\frac{\sigma_0}{\sigma}d_W^2(H_\infty,G).
\end{equation}
Indeed, writing $x=G^{-1}-\mu$ and using antisymmetry gives
$d_W^2(F,G)=A_F-2\langle s_F,x\rangle$, while
$d_W^2(H_\infty,G)=2\sigma^2-2\sigma\langle s_F,x\rangle/\sigma_0$.
Thus, on the fixed-moment symmetric class, the budget remaining after
$\epsilon_0$ is a rescaled squared distance from $H_\infty$.

By Lemma~\ref{lem:sym-feasibility}, only budgets
$\epsilon>\epsilon_0$ require further optimization.
If instead $F=\delta_a$, \notationedit{the point-mass distribution at $a$}, then
$d_W^2(F,G)=(\mu-a)^2+\sigma^2$ for every
$G\in\mathcal M^s(\mu,\sigma)$. In that case the uncertainty set is
empty below this threshold and equals \notationedit{the full class of symmetric
distributions with the prescribed mean and variance}
at or above it, so Lemma~\ref{le.5} applies without a
correlation or multiplier calculation.

Let $g_{\lambda}(t)= \bar{g}(t) + \lambda [\int_{1-t}^{1} F^{-1}(s) \mathrm{d}s+  \int_{t}^{1} F^{-1}(s) \mathrm{d}s-\mu_{F}]$ for $t\in[0,1]$ and $\lambda\geq0$. Whenever $b_\lambda>0$, define $H_\lambda$ by the quantile
\begin{align}\label{(5)}
 h_\lambda(t) = \mu + \sigma \frac{(g_\lambda^{\ast})'(1-t) - a_\lambda}{b_\lambda},
\end{align}
where $a_\lambda=\mathbb E[(g_\lambda^*)'(U)]=0$ and
$b_\lambda=\notationedit{\|(g_\lambda^*)'\|}$, as justified by
Lemma~\ref{lem:sym-envelope-regularity}. In particular,
$b_0=\notationedit{\|(\bar g^*)'\|}$, and $H_0$ is defined only when $b_0>0$.
If $b_0=0$, then $\bar g^*\equiv0$ and $\bar g\leq0$;
this does not imply $\bar g\equiv0$.

Both $g_\lambda$ and its concave envelope are symmetric about $1/2$.
Hence $(g_{\lambda}^{\ast})'(t)+(g_{\lambda}^{\ast})'(1-t)=0$ almost
everywhere, and $h_\lambda(1-t)+h_\lambda(t)=2\mu$ almost everywhere.
The corresponding left and right quantiles satisfy
\eqref{eq:quantile-symmetry} at every interior probability level.

Throughout this section, we impose the following assumption.
\begin{assumption}\label{assum.1}
Suppose that $g\in\mathcal G_2$ and that $(g_{\lambda}^{\ast})'$
is nonconstant for every $\lambda>0$.
\end{assumption}
\begin{remark}[Nondegeneracy]\label{rem:nondegeneracy}
For nondegenerate $F$, the condition $\bar g\geq0$ implies the
nondegeneracy required in Assumption~\ref{assum.1}. Indeed,
$\lambda[\int_{1/2}^1F^{-1}(u)\,du-\int_0^{1/2}F^{-1}(u)\,du]>0$
for every $\lambda>0$, so $g_\lambda^*(1/2)>0$.
Since $g_\lambda^*$ vanishes at both endpoints, its derivative cannot
be constant. The ES, RVaR and upper-tail VaR levels used in
Section~\ref{sec5}, as well as Gini and mean-median deviation,
satisfy this condition.
The assumption does not follow from $g\in\mathcal G_2$ alone.
For example, if $g(t)=-t(1-t)$ and $F=U[-\sqrt3,\sqrt3]$, then
$g_\lambda(t)=(2\sqrt3\lambda-1)t(1-t)$ has zero concave envelope
for $0<\lambda\leq1/(2\sqrt3)$. Nondegeneracy therefore needs to
be verified for $-g$ when using the best-case reduction.
\end{remark}
The conditions defining $\mathcal G_2$ also control the
symmetrized envelopes.
\begin{lemma}[Regularity of the symmetrized envelopes]
\label{lem:sym-envelope-regularity}
Let $g\in\mathcal G_2$ and $F\in\mathcal M^2$.
Then $\bar g^*$ and $g_\lambda^*$ are absolutely continuous on
$[0,1]$, vanish at both endpoints, and have square-integrable
derivatives for every $\lambda\geq0$. In particular,
\begin{equation}\label{eq:sym-envelope-means}
\int_0^1(\bar g^*)'(t)\,\mathrm dt=0,
\qquad
a_\lambda=\int_0^1(g_\lambda^*)'(t)\,\mathrm dt=0.
\end{equation}
\end{lemma}

Since $g\in\mathcal G_2$, the riskmetric $\rho_g$ is finite on
$\mathcal M^2$. For any $G\in\mathcal M_\epsilon^s(\mu,\sigma)$,
Proposition~\ref{prop.1} and
Lemma~\ref{lem:sym-envelope-regularity} yield
\begin{align*}
\rho_g(G)
&=\mu g(1)+\rho_{\bar g}(G)\\
&\leq\mu g(1)+\rho_{\bar g^*}(G)\\
&=\mu g(1)+\int_0^1(\bar g^*)'(1-u)
\bigl(G^{-1}(u)-\mu\bigr)\,\mathrm du\\
&\leq\mu g(1)+\sigma
\left(\int_0^1\bigl[(\bar g^*)'(t)\bigr]^2\,\mathrm dt\right)^{1/2}
<\infty.
\end{align*}
Here the centering by $\mu$ follows from
\eqref{eq:sym-envelope-means}, and the last inequality is the
Cauchy--Schwarz inequality.
For the fixed probability levels considered in this paper,
the distortions associated with $\VaR$, $\ES$, $\RVaR$, Gini deviation,
Gini's premium principle, and mean-median deviation belong to
$\mathcal G_2$: their concave and convex envelopes are Lipschitz
continuous and hence have square-integrable derivatives.
\subsection{Sharp bounds, relaxation, and attainment}\label{subsec:sym-bounds}

To derive sharp upper bounds, define $$\hat{g}(t)=\max\{g(t-),~g(t)~\text{and}~g(t+)\}$$ for $t\in(0,1)$, and $\hat{g}(t)=g(t)$ for $t=0,1$, which is the upper semicontinuous modification of $g$ on the interior of the unit interval.

\begin{lemma}[Symmetric approximation]\label{lem:sym-usc-approximation}
For $g\in\mathcal G_2$ and $G\in\mathcal M^s(\mu,\sigma)$, there
exist $G_n\in\mathcal M^s(\mu,\sigma)$ such that
\[
d_W(G_n,G)\longrightarrow0,\qquad
\rho_{\bar g}(G_n)\longrightarrow\rho_{\hat{\bar g}}(G).
\]
\end{lemma}

\begin{lemma}\label{le.5}
For $\mu\in\mathbb R$, $\sigma>0$, and $g\in\mathcal G_2$,
\[
\sup_{G\in\mathcal M^s(\mu,\sigma)}\rho_g(G)
=\mu g(1)+\sigma b_0.
\]
If $b_0>0$, then $\rho_{\hat{\bar g}}(H_0)=\sigma b_0$,
and $H_0$ is the unique maximizer when $\bar g=\hat{\bar g}$.
If $b_0=0$, the supremum equals $\mu g(1)$ and need not be attained;
if in addition $\bar g\equiv0$, every distribution in
$\mathcal M^s(\mu,\sigma)$ is a maximizer.
\end{lemma}

\begin{lemma}\label{le.4}
Suppose that $g\in\mathcal G_2$ and $\bar g=\hat{\bar g}$.
For every $\lambda>0$ under Assumption~\ref{assum.1},
\[
\sup_{G\in\mathcal M^s(\mu,\sigma)}\rho_{g_\lambda}(G)
=\rho_{g_\lambda}(H_\lambda)=\sigma b_\lambda,
\]
with unique maximizer $H_\lambda$ defined by \eqref{(5)}.
The same statement holds at $\lambda=0$ if $b_0>0$.
\end{lemma}

\begin{lemma}\label{le.3}
Under Assumption~\ref{assum.1}, the function
\[
c(\lambda)=\mathrm{Corr}\bigl(F^{-1}(U),(g_\lambda^*)'(1-U)\bigr),
\qquad \lambda>0,
\]
is continuous and nondecreasing, with
$\lim_{\lambda\to\infty}c(\lambda)=\sigma_0/\sigma_F$.
Consequently, the right limit
$c_0:=\lim_{\lambda\downarrow0}c(\lambda)$ exists.
If $b_0>0$, this limit equals
$\mathrm{Corr}(F^{-1}(U),(\bar g^*)'(1-U))$ and defines a continuous
extension to zero. If $b_0=0$, no correlation at zero is defined.
\end{lemma}

Define the upper budget threshold by
\begin{equation}\label{eq:symmetric-upper-budget}
\epsilon_1=A_F-2\sigma\sigma_Fc_0\geq\epsilon_0.
\end{equation}
Then $d_W^2(F,H_\lambda)=A_F-2\sigma\sigma_Fc(\lambda)$
is nonincreasing for $\lambda>0$, with limits $\epsilon_1$ at zero
and $\epsilon_0$ at infinity. Hence, if $b_0>0$, then
$\epsilon_1=d_W^2(F,H_0)$.

\begin{theorem}\label{th.3}
Suppose that Assumption~\ref{assum.1} holds and
$\bar g=\hat{\bar g}$. Then:
\begin{enumerate}[(i)]
\item If $\epsilon_0<\epsilon<\epsilon_1$, then
\begin{align*}
\sup_{G\in \mathcal{M}_{\epsilon}^{s} (\mu,\sigma)}\rho_{g}(G)=\rho_{g}(H_{\lambda_{\epsilon}}),
\end{align*}
where $H_{\lambda_{\epsilon}}$ is the unique worst-case distribution and $\lambda_{\epsilon}>0$ satisfies $d_W(F,H_{\lambda_{\epsilon}})=\sqrt{\epsilon}$.
\item If $\epsilon\geq\epsilon_1$ and $b_0>0$, the worst-case value
is $\mu g(1)+\sigma b_0$, with unique maximizer $H_0$.
\item If $\epsilon\geq\epsilon_1$ and $b_0=0$, the supremum is
$\mu g(1)$ and need not be attained.
If $\bar g\equiv0$, every feasible distribution is a maximizer.
\end{enumerate}
\end{theorem}

With $\epsilon=\infty$, Theorem~\ref{th.3} extends \notationedit{the bounds under
moment and symmetry constraints} in \citet[Theorem~2.2]{shao2023distortion},
\citet[Corollary~3]{shao2024extreme} and \citet[Theorem~1]{li2018worst}
to distortion functions of bounded variation satisfying its assumptions.
With an active Wasserstein constraint, it adds symmetry to the settings
of \citet[Theorem~2]{bernard2024robust} and
\citet[Theorem~1]{liu2025robust}.
%\rev{Theorem~\ref{th.3.1} removes the upper semicontinuity assumption on the symmetrized distortion.}

The next theorem also covers symmetrized distortions that are not upper
semicontinuous. Its approximation argument preserves symmetry and the exact
moments, and applies whenever the budget is strictly above the feasibility
threshold.

\begin{theorem}\label{th.3.1}
Suppose that Assumption~\ref{assum.1} holds.
Then:
\begin{enumerate}[(i)]
\item If $\epsilon_0<\epsilon<\epsilon_1$, then
\begin{align*}
\sup_{G\in \mathcal{M}_{\epsilon}^{s} (\mu,\sigma)}\rho_{g}(G)=\mu g(1)+\sup_{G\in \mathcal{M}_{\epsilon}^{s} (\mu,\sigma)}\rho_{\hat{\bar{g}}}(G)=\mu g(1)+\rho_{\hat{\bar{g}}}(H_{\lambda_{\epsilon}}),
\end{align*}
with $\lambda_{\epsilon}>0$ satisfying $d_W(F,H_{\lambda_{\epsilon}})=\sqrt{\epsilon}$.
\item If $\epsilon\geq\epsilon_1$, then
\[
\sup_{G\in\mathcal M_\epsilon^s(\mu,\sigma)}\rho_g(G)
=\mu g(1)+\sigma b_0.
\]
For $b_0>0$, the additional term equals
$\rho_{\hat{\bar g}}(H_0)$; for $b_0=0$, it is zero and $H_0$
is not defined. Neither case asserts attainment for $\rho_g$.
 \end{enumerate}
\end{theorem}
Subject to its assumptions, Theorem~\ref{th.3.1} covers the distortions
in $\mathcal G_2$ associated with VaR, GlueVaR and interquantile
differences; see \citet[Section~3]{liu2025robust}. It gives the supremum
without guaranteeing that a maximizing distribution exists.
The VaR example below illustrates this distinction.

\begin{remark}[Order of relaxation and the feasibility boundary]
\label{rem:relaxation-order}
The relaxed distortion in Theorem~\ref{th.3.1} is
$\widehat{\bar g}$: symmetrization is performed before upper
semicontinuous modification. These operations need not commute.
For $g(t)=\mathbf1_{(1/2,1]}(t)$, for example,
\[
\bar g(t)=-\tfrac12\mathbf1_{\{1/2\}}(t),\qquad
\widehat{\bar g}\equiv0,\qquad
\overline{\hat g}(t)=\tfrac12\mathbf1_{\{1/2\}}(t).
\]
The probability level $1/2$ therefore requires separate attention in
the VaR analysis. At $\epsilon=\epsilon_0$, the feasible
set is the singleton $\{H_\infty\}$, so the value is
$\rho_g(H_\infty)$ itself. The relaxation identity cannot in general
be extended to this boundary: the approximation requires strict budget
slack, and the original and relaxed risks at $H_\infty$ can differ.
\end{remark}

\subsection{Explicit bounds for concave distortions}\label{subsec:sym-concave}

If $g$ is concave, Theorem~\ref{th.3} yields an explicit expression for the Wasserstein multiplier.
\begin{proposition}\label{th.1}
Let $g\in\mathcal G_2$ be concave and $\epsilon>\epsilon_0$.
\begin{enumerate}[(i)]
\item If $\epsilon_0<\epsilon<\epsilon_1,$
then
\begin{align*}
\sup_{G\in \mathcal{M}_{\epsilon}^{s} (\mu,\sigma)}\rho_{g}(G)=\rho_{g}(H_{\lambda_{\epsilon}}),
\end{align*}
where $H_{\lambda_{\epsilon}}$ is the unique worst-case distribution and \begin{align}\label{e1}
\lambda_\epsilon
=\frac{1}{2\sigma_0^2}\left[
C_{\bar{g},F}+C_{\epsilon,F}\sqrt{\frac{\sigma_0^2V_{\bar{g}}-C_{\bar{g},F}^2}{\sigma^2\sigma_0^2-C_{\epsilon,F}^2}}
\right],\end{align}
with
$C_{\bar{g},F}=\mathrm{Cov}(F^{-1}(U),\bar{g}'(U))\leq0,~
V_{\bar{g}}=\mathrm{Var}(\bar {g}'(U)),$
$C_{\epsilon,F}=\frac{(\mu-\mu_F)^2+\sigma^2+\sigma_F^2-\epsilon}{2}\ge0.$
\item  Suppose $\epsilon\geq \epsilon_1$.
If $V_{\bar g}>0$, the value is $\mu g(1)+\sigma\sqrt{V_{\bar g}}$
and the unique maximizer is $H_0$. If $V_{\bar g}=0$, concavity implies
$g(t)=tg(1)$; then $\epsilon_1=\epsilon_0$ and every feasible
distribution has the same value $\mu g(1)$.
\end{enumerate}
\end{proposition}

When $\epsilon=\infty$, the Wasserstein constraint is removed. If, in addition, $g(t)$ is increasing with $g(1)=1$, Proposition~\ref{th.1} recovers the bounds for classical distortion risk measures under moment and symmetry constraints in Proposition~2.2 of \cite{shao2023distortion}.

Proposition~\ref{th.1} gives the following formulas for ES, GD and MMD.
Their counterparts without symmetry were obtained by
\cite{bernard2024robust} for ES and by \cite{liu2025robust} for GD and MMD.

\begin{corollary}\label{ex.1} Suppose $\mu\in\R, \sigma>0$ and $\epsilon_0<\epsilon<\epsilon_1$.
\begin{enumerate}[(i)]
\item Expected Shortfall (ES). For $\alpha>\frac{1}{2}$, the worst-case $\ES$ is given by
\begin{align*}
\sup_{G\in \mathcal{M}_{\epsilon}^{s} (\mu,\sigma)}\mathrm{ES}_{\alpha}(G)&=\mu+\sigma\left[\frac{1}{2(1-\alpha)\varpi}-\frac{2\lambda_{\epsilon}C_{\bar{g},F}}{\varpi}\right],
\end{align*}
where $\varpi=
\sqrt{\frac{1}{2(1-\alpha)}+4\lambda_\epsilon^2\sigma_0^2-4\lambda_\epsilon C_{\bar{g},F}}$, and the worst-case quantile is given by $$h_{\lambda_{\epsilon}}(t)=\mu+\sigma\frac{\frac{{\mathbf{1}_{(\alpha,1]}(t)}-\mathbf{1}_{[0,1-\alpha]}(t)}{2(1-\alpha)}+\lambda_{\epsilon}(F^{-1}(t)-{F^{-1+}(1-t)})}{\varpi},~t\in (0,1),$$ where $\lambda_\epsilon$ is given by \eqref{e1}
with $C_{\bar{g},F}=\frac{\mu_{F}-\alpha\mathrm{ES}_{1-\alpha}(F)-(1-\alpha)\mathrm{ES}_{\alpha}(F)}{2(1-\alpha)}\leq0$, $V_{\bar{g}}=\frac{1}{2(1-\alpha)}$ and $C_{\epsilon,F}=\frac{(\mu-\mu_{F})^{2}+\sigma^{2}+\sigma_{F}^{2}-\epsilon}{2}$.
\item  Gini deviation (GD). The worst-case GD is given by
\begin{align*}
\sup_{G\in \mathcal{M}_{\epsilon}^{s} (\mu,\sigma)}\mathrm{GD}(G)=\frac{\sigma}{\varpi}\left[\frac{1}{3}-2\lambda_{\epsilon}C_{\bar{g},F}\right],
\end{align*}
where $\varpi=\sqrt{1/3+4\lambda_{\epsilon}^{2}\sigma_{0}^{2}-4\lambda_{\epsilon}C_{\bar{g},F}}$, and the worst-case quantile is given by
$$h_{\lambda_{\epsilon}}(t)=\mu+\sigma\frac{2t-1+\lambda_{\epsilon}( F^{-1}(t)-{F^{-1+}(1-t)})}{\varpi}, ~t\in (0,1),$$ where $\lambda_\epsilon$ is given by \eqref{e1}
with $C_{\bar{g},F}=\int_{0}^{1}(1-2t)F^{-1}(t)\mathrm{d}t\leq0$. The other constants are $V_{\bar{g}}=\frac{1}{3}$ and $C_{\epsilon,F}=\frac{(\mu-\mu_{F})^{2}+\sigma^{2}+\sigma_{F}^{2}-\epsilon}{2}$.
\item Mean-median deviation (MMD). The worst-case MMD is given by
\begin{align*}
\sup_{G\in \mathcal{M}_{\epsilon}^{s} (\mu,\sigma)}\mathrm{MMD}(G)=\frac{\sigma}{\varpi}\left[1-2\lambda_{\epsilon}C_{\bar{g},F}\right],
\end{align*}
where $\varpi=\sqrt{1+4\lambda_\epsilon^2\sigma_0^2-4\lambda_\epsilon C_{\bar{g},F}}$, and the worst-case quantile is given by
$$h_{\lambda_{\epsilon}}(t)=\mu+\sigma\frac{\mathbf{1}_{(\frac{1}{2},1]}(t)-\mathbf{1}_{[0,\frac{1}{2}]}(t)+\lambda_{\epsilon} (F^{-1}(t)-{F^{-1+}(1-t)})}{\varpi}, ~t\in (0,1),$$
 where $\lambda_\epsilon$ is given by  \eqref{e1}
with $C_{\bar{g},F}=\int_{0}^{\frac{1}{2}}F^{-1}(t)\mathrm{d}t-\int_{1/2}^{1}F^{-1}(t)\mathrm{d}t$, $V_{\bar{g}}=1$ and $C_{\epsilon,F}=\frac{(\mu-\mu_{F})^{2}+\sigma^{2}+\sigma_{F}^{2}-\epsilon}{2}$.
\end{enumerate}
\end{corollary}

As $\epsilon\rightarrow\infty$, the Wasserstein constraint becomes inactive. Corollary~\ref{ex.1}(i) then gives
\[
\sup_{G\in \mathcal{M}^{s} (\mu,\sigma)}\mathrm{ES}_{\alpha}(G)=\mu+\frac{\sigma}{\sqrt{2(1-\alpha)}}
\]
and the worst-case quantile is given by
\begin{align*}
h_{0}(t)=
\begin{cases}
\mu+\frac{\sigma}{\sqrt{2(1-\alpha)}},~&t\in(\alpha,1],\\
\mu,~&t\in(1-\alpha,\alpha],\\
\mu-\frac{\sigma}{\sqrt{2(1-\alpha)}},~&t\in[0,1-\alpha].
\end{cases}
\end{align*}
This agrees with Lemma~2.6 of \cite{li2018worst}.

\subsection{Range Value-at-Risk}\label{subsec:sym-rvar}
We now apply Theorem~\ref{th.3} to RVaR. Explicit bounds are known
under moment and Wasserstein constraints
\citep{bernard2024robust,liu2025robust}, and under moment and symmetry
constraints \citep{li2018worst,shao2023distortion}.
The construction below imposes all three restrictions.

For $\frac{1}{2}\le\alpha<\beta<1$ and $\lambda\ge0$, define
 $$r_\lambda(t)=\lambda\left[
\int_{1-t}^1 F^{-1}(s)\mathrm{d} s+\int_t^1 F^{-1}(s)\mathrm{d} s-\mu_F\right],~t\in[0,1].$$
On $[0,1-\alpha]$, the function $g_\lambda$ consists of two concave branches:
\[
g_\lambda(t)=
\begin{cases}
r_\lambda(t),&0\le t\le1-\beta,\\
r_\lambda(t)+\dfrac{t-(1-\beta)}{2(\beta-\alpha)},
&1-\beta\le t\le1-\alpha.
\end{cases}
\]
For a fixed slope $m\in\mathbb R$, let
 \begin{align*}
A_\lambda(m)&=\max_{0\le t\le1-\beta}
\{r_\lambda(t)-mt\},\\
B_\lambda(m)&=\max_{1-\beta\le t\le1-\alpha}
\left\{\frac{t-(1-\beta)}{2(\beta-\alpha)}
+r_\lambda(t)-mt
\right\}.
\end{align*}
These are the smallest intercepts of affine majorants with slope $m$
for the first and second branches, respectively. A common upper supporting
line is obtained when $A_\lambda(m)=B_\lambda(m)$. Set
\[
m_\lambda=\min\{m\in\mathbb R:A_\lambda(m)=B_\lambda(m)\},
\qquad
\ell_\lambda(t)=m_\lambda t+A_\lambda(m_\lambda).
\]
We define $u_\lambda$ and $v_\lambda$ as the leftmost and rightmost
contact points of this line with the respective branches:
\begin{align*}
u_\lambda&=\min\{t\in[0,1-\beta]:g_\lambda(t)=\ell_\lambda(t)\},\\
v_\lambda&=\max\{t\in[1-\beta,1-\alpha]:g_\lambda(t)=\ell_\lambda(t)\}.
\end{align*}
This convention also covers nonunique contact points. The concave envelope
$g_\lambda^\ast$ coincides with $\ell_\lambda$ on $[u_\lambda,v_\lambda]$;
its slope there is $m_\lambda$ when $u_\lambda<v_\lambda$.

Theorem~\ref{th.3} yields the following RVaR bound. %{ It is better to give all results for RVaR}.
\begin{corollary}[RVaR]\label{ex.5} Suppose $\epsilon_0<\epsilon<\epsilon_1$.
 If $\frac{1}{2}\leq\alpha<\beta<1$, then
\begin{align*}
&\sup_{G\in \mathcal{M}_{\epsilon}^{s} (\mu,\sigma)}\text{RVaR}_{\alpha,\beta}(G)\\
&=\mu+\frac{\sigma}{(\beta-\alpha)b_\lambda}\Biggl\{
\frac{1-\alpha-v_\lambda}{2(\beta-\alpha)}
+m_\lambda(\beta-1+v_\lambda)
+\lambda
\int_{v_\lambda}^{1-\alpha}[F^{-1}(1-s)-F^{-1}(s)]\mathrm{d} s
\Biggr\},
\end{align*}
and the worst-case quantile is $h_{\lambda_{\epsilon}}$ from
 \eqref{(5)} with
$$(g_{\lambda}^*)'(1-t)
=\begin{cases}
\lambda\bigl[F^{-1}(t)-F^{-1+}(1-t)\bigr],
&0< t\leq u_\lambda,\\
-m_\lambda,
&u_\lambda<t\leq v_\lambda,\\
-\dfrac1{2(\beta-\alpha)}
+\lambda\bigl[F^{-1}(t)-F^{-1+}(1-t)\bigr],
&v_\lambda<t\leq1-\alpha,\\
\lambda\bigl[F^{-1}(t)-F^{-1+}(1-t)\bigr],
&1-\alpha<t\leq\alpha,\\
\dfrac1{2(\beta-\alpha)}
+\lambda\bigl[F^{-1}(t)-F^{-1+}(1-t)\bigr],
&\alpha<t\leq1-v_\lambda,\\
m_\lambda,
&1-v_\lambda<t\leq1-u_\lambda,\\
\lambda\bigl[F^{-1}(t)-{F^{-1+}(1-t)}\bigr],
&1-u_\lambda<t<1,
\end{cases}
$$
where $\lambda_\epsilon>0$ satisfies $d_W(F,H_{\lambda_{\epsilon}})=\sqrt{\epsilon}$.
All quantities indexed by $\lambda$ in this corollary are evaluated at
$\lambda=\lambda_\epsilon$.
\end{corollary}
We state the formula for $\frac12\leq\alpha<\beta<1$, the range
most relevant to upper-tail risk. Theorem~\ref{th.3} also applies when
$0<\alpha<\beta<1$, but the explicit expressions then require additional
cases.

If $\epsilon=\infty$, then the result in Corollary \ref{ex.5} reduces to
\begin{align*}
\sup_{G\in \mathcal{M}^{s} (\mu,\sigma)}\text{RVaR}_{\alpha,\beta}(G)
=&\mu+\frac{\sigma}{\sqrt{2(1-\alpha)}},
\end{align*}
and the corresponding worst-case quantile is
$$\mu+\frac{\sigma}{\sqrt{2(1-\alpha)}}(\mathbf{1}_{(\alpha,1]}(u)-\mathbf{1}_{[0,1-\alpha]}(u)), ~u\in[0,1].$$
This recovers Theorem~2.2 of \cite{li2018worst} and the corresponding entry in Table~1 of \cite{shao2023distortion}.

\subsection{VaR and nonattainment}\label{subsec:sym-var}

For $g(t)=\mathbf1_{(1-\alpha,1]}(t)$, $\alpha\in(0,1)$,
the symmetrized distortion $\bar g$ is not upper semicontinuous.
We therefore apply Theorem~\ref{th.3.1} to $\VaR_\alpha$. Write
\[
R_F(t)=\int_{1-t}^1F^{-1}(u)\,\mathrm du
      +\int_t^1F^{-1}(u)\,\mathrm du-\mu_F,
\qquad g_\lambda=\bar g+\lambda R_F.
\]
The contact points below are defined by extremal chord slopes, so
no continuity of the reference quantile is required.
For $\alpha>1/2$, set $a=1-\alpha$ and let $t_{\alpha,\lambda}$
be the smallest minimizer of
\begin{equation}\label{eq:var-upper-contact}
\frac{1/2+\lambda[R_F(a)-R_F(t)]}{a-t},\qquad 0\leq t<a.
\end{equation}
Denote the minimum by $m_{\alpha,\lambda}^+$.
For $\alpha<1/2$, set
$\Delta_{\alpha,\lambda}=\lambda[R_F(1/2)-R_F(\alpha)]$.
When $\Delta_{\alpha,\lambda}>1/2$, let
$\tilde t_{\alpha,\lambda}$ be the smallest maximizer of
\begin{equation}\label{eq:var-lower-contact}
\frac{-1/2+\lambda[R_F(t)-R_F(\alpha)]}{t-\alpha},
\qquad \alpha<t\leq1/2,
\end{equation}
whose maximum is denoted by $m_{\alpha,\lambda}^->0$.
The extrema are attained: the quotient in
\eqref{eq:var-upper-contact} tends to $+\infty$ as $t\uparrow a$,
and that in \eqref{eq:var-lower-contact} tends to $-\infty$ as
$t\downarrow\alpha$.

\begin{corollary}[VaR]\label{ex.6}
Let $g(t)=\mathbf1_{(1-\alpha,1]}(t)$, where
$\alpha\in(0,1)\setminus\{1/2\}$, and suppose
$\epsilon_0<\epsilon<\epsilon_1$.
Choose $\lambda=\lambda_\epsilon>0$ satisfying
$d_W^2(F,H_\lambda)=\epsilon$.
%If $\alpha<1/2$, then necessarily $\Delta_{\alpha,\lambda}>1/2$.
The worst-case value is
\begin{equation}\label{eq:var-value}
\sup_{G\in\mathcal M_\epsilon^s(\mu,\sigma)}\VaR_\alpha(G)
=\mu+\frac{\sigma}{b_\lambda}
\begin{cases}
 m_{\alpha,\lambda}^+,&\alpha>1/2,\\
 -m_{\alpha,\lambda}^-,&\alpha<1/2,
\end{cases}
\end{equation}
and the supremum is not attained.
Here $b_\lambda=\notationedit{\|(g_\lambda^*)'\|}$, and $H_\lambda$ has the
quantile in \eqref{(5)}, obtained by differentiating
\eqref{eq:var-upper-envelope} or \eqref{eq:var-lower-envelope}
in the proof.
\end{corollary}

The right-quantile risk $\VaR_\alpha^+$ has the upper semicontinuous
distortion $\mathbf1_{[1-\alpha,1]}(t)$. Its symmetrization equals
$\hat{\bar g}$. Under the conditions of Corollary~\ref{ex.6},
Theorems~\ref{th.3} and~\ref{th.3.1} therefore give the same supremum
for $\VaR_\alpha$ and $\VaR_\alpha^+$, and identify
$H_{\lambda_\epsilon}$ as the unique maximizer of $\VaR_\alpha^+$.
The strict inequalities established in the proof show that its left
$\alpha$-quantile is below this common supremum, for both
$\alpha<1/2$ and $\alpha>1/2$.
For $\alpha>1/2$, nonattainment also holds when
$\epsilon\geq\epsilon_1$ and $\epsilon>\epsilon_0$; hence the left-VaR
supremum is unattained for every budget strictly above the feasibility
threshold. For related nonattainment results without symmetry,
see \cite{bernard2024robust,Pesenti2025robust,liu2025robust}. With the Wasserstein constraint removed, Lemma~\ref{le.5}
gives, for $\alpha>1/2$,
\begin{align*}
&\sup_{G\in \mathcal{M}^{s} (\mu,\sigma)}\mathrm{VaR}_{\alpha}(G)=
\mu +\frac{\sigma}{\sqrt{2(1-\alpha)}},
\end{align*}
in agreement with Lemma~2.6 of \cite{li2018worst}.

\section{Robust distortion riskmetrics under symmetry and unimodality}\label{sec4}
We now impose unimodality in addition to symmetry.
Throughout this section, $F\in\mathcal M^2$ and $g$ is absolutely
continuous with $\int_0^1(g'(t))^2\d t<\infty$.

\begin{lemma}[Geometry of the projection cone]\label{lem:us-cone}
The set $\mathcal M^{us,-1}$ is a closed convex cone in $L^2(0,1)$.
Let $\mathcal C$ be its centered subcone and write
$A_f=\int_0^1 f(u)\,du$ and
$f^{\mathrm a}(u)=[f(u)-f(1-u)]/2$ for $f\in L^2(0,1)$.
If $P_{us}$ and $P_{\mathcal C}$ denote the respective projections, then
\begin{equation}\label{eq:generalidentity}
P_{us}f=A_f+P_{\mathcal C}f^{\mathrm a},\qquad
P_{us}f^{\mathrm a}=P_{\mathcal C}f^{\mathrm a},\qquad
\int_0^1P_{us}f(u)\,du=A_f.
\end{equation}
\end{lemma}
Lemma~\ref{lem:us-cone} and the Hilbert-space projection theorem
\cite[Theorem~2.1]{brunk1965conditional} ensure that, for $\lambda\geq0$,
the $L^{2}$-projection of $k_{\lambda}(u)=\bar{g}'(1-u)+\lambda F^{-1}(u)$ onto $\mathcal{M}^{us,-1}$ exists and is unique. Denote it by
\begin{align*}
k_{\lambda,us}^{\uparrow}=\arg\min_{k\in\mathcal{M}^{us,-1}}\| k_{\lambda}-k\|^{2},
\end{align*}
For $\lambda>0$, define the standardized projection
\begin{align}\label{Eq:us}
h_{\lambda,us}^{\uparrow}(u) = \mu+\sigma\frac{k_{\lambda,us}^{\uparrow}(u)-a_{\lambda,us}}{b_{\lambda,us}},
\end{align}
where $a_{\lambda,us}=\mathbb{E}(k_{\lambda,us}^{\uparrow}(U))$ and $b_{\lambda,us}=\sqrt{\mathrm{Var}(k_{\lambda,us}^{\uparrow}(U))}$. The normalization in \eqref{Eq:us} is well defined only if $b_{\lambda,us}>0$. Whenever this condition holds, $h_{\lambda,us}^{\uparrow}$ is a quantile function; denote its distribution by $H_{\lambda,us}^{\uparrow}$. We also define
under the nondegeneracy assumptions of Proposition~\ref{le.6},
\[
{c_{us}(\lambda)=\mathrm{Corr}(F^{-1}(U),k_{\lambda,us}^{\uparrow}(U)),
\quad\lambda>0,\qquad c_1=\lim_{\lambda\downarrow0}c_{us}(\lambda).}
\]
Let $F^{-1,\uparrow}$ denote the $L^{2}$-projection of $F^{-1}$ onto $\mathcal{M}^{us,-1}$, and set $\bar{c}_{0}=\mathrm{Corr}(F^{-1}(U),F^{-1,\uparrow}(U))$ whenever $F^{-1,\uparrow}$ is nonconstant. Define the two Wasserstein budget thresholds $\overline{\epsilon}_0=(\mu_{F}-\mu)^2+(\sigma_{F}-\sigma)^2+2\sigma_{F}\sigma(1-\bar{c}_{0})$ and $\overline{\epsilon}_1=(\mu_{F}-\mu)^2+(\sigma_{F}-\sigma)^2+2\sigma_{F}\sigma(1-c_{1})$. Henceforth, we retain $\epsilon_0,\epsilon_1$ for the symmetry-only thresholds
and use $\overline{\epsilon}_0,\overline{\epsilon}_1$ for the joint
symmetry and unimodality constraints. Inclusion of the uncertainty sets
implies $\overline{\epsilon}_0\geq\epsilon_0$; the two saturation
thresholds are defined by their respective auxiliary optimizers.

For the zero-multiplier case, write
$\bar{g}'^{\uparrow}:=k_{0,us}^{\uparrow}$, the projection of
$u\mapsto\bar{g}'(1-u)$ onto $\mathcal{M}^{us,-1}$.
The normalization in \eqref{Eq:us} is also used at $\lambda=0$
whenever $b_{0,us}>0$.

\begin{remark}[A sufficient condition for nondegeneracy]\label{rem:us-nondegeneracy}
If $F$ is nondegenerate, $F^{-1,\uparrow}$ is nonconstant.
If, in addition, $\bar g\geq0$, then $k_{\lambda,us}^{\uparrow}$
is nonconstant for every $\lambda>0$. Indeed, $v(u)=2u-1$ belongs
to $\mathcal C$ and $\langle F^{-1},v\rangle>0$. Integration by parts gives
\[
\langle k_\lambda^{\mathrm a},v\rangle
=2\int_0^1\bar g(u)\,du+\lambda\langle F^{-1},v\rangle>0.
\]
A function with positive inner product with an element of $\mathcal C$
cannot have zero projection onto that cone. The same argument applied
to $F^{-1}$ proves the first assertion. This condition covers ES, RVaR,
Gini deviation and mean-median deviation.
\end{remark}
The projection of $F^{-1}$ determines the minimum feasible budget.
\begin{lemma}\label{le.6a}
Suppose $F\in\mathcal M^2$ and $F^{-1,\uparrow}$ is nonconstant. If $\epsilon<\overline{\epsilon}_0$, then
$\mathcal{M}^{us,-1}_{\epsilon}(\mu,\sigma)=\emptyset$. If $\epsilon=\overline{\epsilon}_0$, then $\mathcal{M}^{us,-1}_{\overline{\epsilon}_0}(\mu,\sigma)=\left\{\mu+\sigma\frac{F^{-1,\uparrow}-\mathbb{E}(F^{-1,\uparrow}(U))}{\sqrt{\mathrm{Var}(F^{-1,\uparrow}(U))}}\right\}$.
\end{lemma}

It remains to optimize over budgets $\epsilon>\overline{\epsilon}_0$.
The following auxiliary problem determines the candidate quantiles.

\begin{proposition}\label{le.6}Suppose $F\in\mathcal M^2$, and let $g\in\mathcal G$ be
absolutely continuous with $g'\in L^2(0,1)$.
For any $\lambda\geq0$ with $b_{\lambda,us}>0$, the optimization problem
\begin{align*}
\sup_{h\in\mathcal{M}^{us,-1}(\mu,\sigma)}\langle k_{\lambda}(u),h(u)\rangle
\end{align*}
has the unique solution $h_{\lambda,us}^{\uparrow}$ given by \eqref{Eq:us}.
%h_{\lambda,us}^{\uparrow}(u)=\mu+\sigma\frac{k_{\lambda,us}^{\uparrow}(u)-a_{\lambda,us}}{b_{\lambda,us}}.
Moreover, if $F^{-1,\uparrow}$ is nonconstant and $k_{\lambda,us}^{\uparrow}$ is nonconstant for every $\lambda>0$, then $\mathrm{Corr}(F^{-1}(U),k_{\lambda,us}^{\uparrow}(U))$ is continuous and nondecreasing in $\lambda$ on $(0,\infty)$, and
\begin{align*}
\lim_{\lambda\rightarrow\infty}\mathrm{Corr}(F^{-1}(U),k_{\lambda,us}^{\uparrow}(U))=\mathrm{Corr}(F^{-1}(U),F^{-1,\uparrow}(U)).
\end{align*}
The right limit $c_1$ therefore exists. If $b_{0,us}>0$,
it equals $\mathrm{Corr}(F^{-1}(U),k_{0,us}^{\uparrow}(U))$.
Consequently, $d_W^2(F,H_{\lambda,us}^{\uparrow})$ is continuous
and nonincreasing for $\lambda>0$, with limits
$\overline{\epsilon}_1$ at zero and $\overline{\epsilon}_0$ at infinity.
 \end{proposition}

Choosing the multiplier to satisfy the active distance constraint
gives the bound over $\mathcal{M}^{us}_{\epsilon}(\mu,\sigma)$.
\begin{theorem}\label{th.4}
Suppose $F\in\mathcal M^2$, and let $g\in\mathcal G$ be absolutely
continuous with $g'\in L^2(0,1)$. Assume also that
$F^{-1,\uparrow}$ is nonconstant and $k_{\lambda,us}^{\uparrow}$
is nonconstant for every $\lambda>0$.
For $\epsilon>\overline{\epsilon}_0$, write
\[
V_{us}(\epsilon)
:=\sup_{G\in\mathcal M^{us}_{\epsilon}(\mu,\sigma)}\rho_g(G),
\]
and, for finite $\epsilon$, set
\[
D_\epsilon
:=\frac{(\mu-\mu_F)^2+\sigma^2+\sigma_F^2-\epsilon}{2}.
\]
Then the following statements hold.
\begin{enumerate}[(i)]
\item If $\overline{\epsilon}_0<\epsilon<\overline{\epsilon}_1$, then
\begin{align}\label{eq:us-sharp-value}
V_{us}(\epsilon)
&=\rho_g(H_{\lambda_\epsilon,us}^{\uparrow})\nonumber\\
&=\mu g(1)+\sigma b_{\lambda_\epsilon,us}
  -\lambda_\epsilon D_\epsilon,
\end{align}
where $H_{\lambda_\epsilon,us}^{\uparrow}$ is the unique worst-case
distribution and $\lambda_\epsilon>0$ satisfies
$d_W(F,H_{\lambda_\epsilon,us}^{\uparrow})=\sqrt\epsilon$.
\item If $\epsilon\geq\overline{\epsilon}_1$, then
\[
V_{us}(\epsilon)=\mu g(1)+\sigma b_{0,us}.
\]
If $b_{0,us}>0$, the unique maximizer is $H_{0,us}^{\uparrow}$.
If $b_{0,us}=0$, the supremum is $\mu g(1)$ and need not be attained.
\end{enumerate}
For every finite $\epsilon>\overline{\epsilon}_0$, the value also
admits the one-dimensional dual representation
\begin{equation}\label{eq:us-dual}
V_{us}(\epsilon)
=\mu g(1)+\inf_{\lambda\geq0}
\left\{\sigma b_{\lambda,us}-\lambda D_\epsilon\right\}.
\end{equation}
The infimum is attained at any multiplier $\lambda_\epsilon$
in part~(i), and at $\lambda=0$ in part~(ii).
\end{theorem}
Theorem~\ref{th.4} adds symmetry to the moment, Wasserstein and
unimodality constraints in \citet[Theorem~3]{liu2025robust}.
Without unimodality, \citet[Theorem~2]{bernard2024robust} treats increasing
distortions under moment and Wasserstein constraints.
For comparison, the RVaR results in \citet[Theorem~2.2]{li2018worst}
and \citet[Theorem~2]{Bernard2020range} impose moment and shape
constraints without a Wasserstein restriction.

The projection argument uses the $L^2$ representation
of the riskmetric through $g'$. ES, RVaR, Gini deviation and
mean-median deviation satisfy the required regularity.
Discontinuous distortions require a different argument and are not
covered by this section.

Without the Wasserstein constraint, Theorem~\ref{th.4}(ii) gives
the following moment-and-shape bound.
\begin{proposition}\label{prop:us-moments}
Let $g\in\mathcal G$ be absolutely continuous with
$g'\in L^2(0,1)$. If $\bar{g}'^{\uparrow}$ is nonconstant, then
\begin{align*}
\sup_{G\in \mathcal{M}^{us} (\mu,\sigma)}\rho_{g}(G)=\rho_{g}(H_{0,us}^{\uparrow}),
\end{align*}
where $H_{0,us}^{\uparrow}$ is the unique worst-case distribution.
\end{proposition}

For Gini's premium principle, the projection is explicit when the
reference is itself symmetric and unimodal. We give this case first,
then approximate the general projection using the step-function
representation of \citet[Proposition~3]{liu2025robust}.

\begin{corollary}[Gini's distortion premium principle]\label{ex.7}
Let $g_r(t)=(1+r)t-rt^2$, with $0<r\leq1$, and suppose that
$F\in\mathcal M^{us}$ with $\sigma_F>0$. Define
\[
C_{r,F}=r\int_0^1(2u-1)F^{-1}(u)\,\mathrm du>0,\qquad
V_r=\frac{r^2}{3},\qquad
D_\epsilon=\frac{(\mu-\mu_F)^2+\sigma^2+\sigma_F^2-\epsilon}{2}.
\]
If $\overline{\epsilon}_0<\epsilon<\overline{\epsilon}_1$, then
\begin{equation}\label{eq:gdp-value}
\sup_{G\in\mathcal M_\epsilon^{us}(\mu,\sigma)}\mathrm{GDP}_r(G)
=\mu+\sigma\frac{V_r+\lambda_\epsilon C_{r,F}}
 {b_{\lambda_\epsilon,us}},
\end{equation}
with unique maximizing quantile
\begin{equation}\label{eq:gdp-quantile}
h_{\lambda_\epsilon,us}^{\uparrow}(u)
=\mu+\sigma\frac{r(2u-1)+\lambda_\epsilon(F^{-1}(u)-\mu_F)}
 {b_{\lambda_\epsilon,us}},\qquad u\in(0,1),
\end{equation}
where $b_{\lambda,us}^2=V_r+2\lambda C_{r,F}+\lambda^2\sigma_F^2$ and
\begin{equation}\label{eq:gdp-multiplier}
\lambda_\epsilon
=\frac{1}{\sigma_F^2}\left[
-C_{r,F}+D_\epsilon
\sqrt{\frac{\sigma_F^2V_r-C_{r,F}^2}
 {\sigma^2\sigma_F^2-D_\epsilon^2}}\right].
\end{equation}
Here $\overline{\epsilon}_0=(\mu-\mu_F)^2+(\sigma-\sigma_F)^2$ and
$\overline{\epsilon}_1=(\mu-\mu_F)^2+\sigma^2+\sigma_F^2
-2\sigma C_{r,F}/\sqrt{V_r}$.
If $C_{r,F}^2=\sigma_F^2V_r$, these thresholds coincide and the
active-budget interval is empty.
For $\epsilon\geq\overline{\epsilon}_1$, setting $\lambda=0$ in
\eqref{eq:gdp-quantile} gives the unique maximizer, with value
$\mu+\sigma r/\sqrt3$.
\end{corollary}
The assumption $F\in\mathcal M^{us}$ makes
$r(2u-1)+\lambda F^{-1}(u)$ an element of the projection cone,
so its projection is itself. This is what gives the explicit formulas.
For a general reference distribution, the projection in
Theorem~\ref{th.4} must still be computed.
When $r=0$, $\mathrm{GDP}_0(G)=\mu$ for every feasible $G$.

For fixed $\lambda$, we approximate $k_\lambda$ by step functions,
for example on uniform partitions of $[0,1]$ into $2n$ subintervals.
The following representation applies to any symmetric partition with
$1/2$ as a node.
\begin{proposition}[Projection of a step function]\label{pro.4}
Let
$
L_n(u)=\sum_{i=1}^{2n}y_{n,i}\mathbf{1}_{(x_{n,i-1},x_{n,i})}(u),
$
where
$
0=x_{n,0}<\cdots<x_{n,n}=\frac12<\cdots<x_{n,2n}=1,
~\text{and}~ x_{n,2n-i}=1-x_{n,i}.
$
Define
$$
a_{L_n}=\int_0^1L_n(u)\,du,
\qquad
\eta_{n,i}=\frac{y_{n,i}-y_{n,2n+1-i}}2,
\quad i=n+1,\ldots,2n.
$$
Then $L_{n,us}^{\uparrow}$ is piecewise linear, has mean
$a_{L_n}$, and satisfies, almost everywhere,
\begin{align}\label{eq:derivative}
\nonumber\bigl(L_{n,us}^{\uparrow}\bigr)'(u)
=\sum_{i=n+1}^{2n}\Big[
&c_{n,i}^-\mathbf{1}_{(x_{n,i-1},\theta_{n,i})}(u)
+c_{n,i}^+\mathbf{1}_{(\theta_{n,i},x_{n,i})}(u)\\
&+c_{n,i}^+\mathbf{1}_{(1-x_{n,i},\,1-\theta_{n,i})}(u)
+c_{n,i}^-\mathbf{1}_{(1-\theta_{n,i},\,1-x_{n,i-1})}(u)
\Big],
\end{align}
and
\begin{equation}\label{eq:endpoint}
L_{n,us}^{\uparrow}(1)
=a_{L_n}+\sum_{i=n+1}^{2n}
\bigl(c_{n,i}^-\ell_{n,i}^-+c_{n,i}^+\ell_{n,i}^+\bigr),
\end{equation}
where
$
\ell_{n,i}^-=\theta_{n,i}-x_{n,i-1},~
\ell_{n,i}^+=x_{n,i}-\theta_{n,i},
$
and the endpoint value is defined by continuous extension.
The admissible parameter set is
$$
\mathcal D_n=\left\{(\theta_n,c_n):
\begin{array}{l}
\theta_{n,i}\in[x_{n,i-1},x_{n,i}],\quad i=n+1,\ldots,2n,\\[2pt]
0\le c_{n,n+1}^-\le c_{n,n+1}^+
\le\cdots\le c_{n,2n}^-\le c_{n,2n}^+
\end{array}
\right\}.
$$
Set
$
e_{n,i}^0=
\sum_{j=n+1}^{i-1}
\bigl(c_{n,j}^-\ell_{n,j}^-+c_{n,j}^+\ell_{n,j}^+\bigr)-\eta_{n,i},
$
$
e_{n,i}^1=e_{n,i}^0+c_{n,i}^-\ell_{n,i}^- ,~
e_{n,i}^2=e_{n,i}^1+c_{n,i}^+\ell_{n,i}^+ ,
$
where an empty sum is interpreted as zero. The optimal parameters solve
\begin{equation}\label{eq:optimization}
(\theta_n^*,c_n^*)\in
\argmin_{(\theta_n,c_n)\in\mathcal D_n}
\sum_{i=n+1}^{2n}\Big\{
\ell_{n,i}^-\big[(e_{n,i}^0)^2+e_{n,i}^0e_{n,i}^1+(e_{n,i}^1)^2\big]
+\ell_{n,i}^+\big[(e_{n,i}^1)^2+e_{n,i}^1e_{n,i}^2+(e_{n,i}^2)^2\big]
\Big\}.
\end{equation}
The projected function is unique, although the minimizing parameters need
not be unique.
\end{proposition}
\begin{proposition}[Fixed-node approximation]\label{prop:fixed-node}
Let $\mathcal C$ be the cone of centered quantiles in
$\mathcal M^{us,-1}$. On nested partitions
$1/2=x_0<\cdots<x_m=1$ with mesh tending to zero, let
$\psi_j(u)=(u-x_j)_+$ for $u\geq1/2$, $j=0,\ldots,m-1$,
and extend each function by $\psi_j(1-u)=-\psi_j(u)$. 
Set $\mathcal C_m=\{\sum_j c_j\psi_j:c_j\geq0\}$.
For $L\in L^2(0,1)$, write $L^{\mathrm a}(u)=[L(u)-L(1-u)]/2$.
Then $p_m=P_{\mathcal C_m}L^{\mathrm a}$ is obtained from the convex
quadratic program
\begin{equation}\label{eq:fixed-node-qp}
\min_{c\geq0}\left\{\frac12c^\top Qc-d^\top c\right\},\qquad
Q_{ij}=\langle\psi_i,\psi_j\rangle,\quad
d_j=\langle L^{\mathrm a},\psi_j\rangle.
\end{equation}
The matrix $Q$ is positive definite. With $z=Qc-d$, the conditions
$c\geq0$, $z\geq0$, and $c^\top z=0$ are necessary and sufficient
for global optimality. Moreover,
$p_m\to P_{\mathcal C}L^{\mathrm a}$ in $L^2$.
If the limiting projection is nonzero, the corresponding standardized
quantiles and their absolutely continuous distortion riskmetrics also
converge.
\end{proposition}
Joint optimization over slopes and breakpoints in
\eqref{eq:optimization} need not be convex. The fixed-node formulation
\eqref{eq:fixed-node-qp} is convex, and its KKT conditions characterize
the projection onto $\mathcal C_m$. If the objective gap in
$\min_{p\in\mathcal C_m}\frac12\notationedit{\|p-L^{\mathrm a}\|}^2$ is at most
$\eta$, the solver error is at most $\sqrt{2\eta}$ in $L^2$.
This bound does not include errors from the grid, input approximation
or numerical integration. Approximating the input by cell averages
avoids evaluating unbounded reference quantiles at $0$ and $1$.
We next relate error in the input to errors in the projection,
the normalized quantile and the risk value. For $f\in L^2(0,1)$, define
$$
f^{\mathrm a}(u)=\frac{f(u)-f(1-u)}2,
$$
$$
a_f=\int_0^1f_{us}^{\uparrow}(u)\,du
=\int_0^1f(u)\,du,
~~~ b_f=\|{f_{us}^{\uparrow}-a_f}\|.
$$
Whenever $b_f>0$, let
$
h_f(u)=\mu+\sigma\frac{f_{us}^{\uparrow}(u)-a_f}{b_f},
$
and denote the distribution with quantile function $h_f$ by $H_f$.
Here $\mu\in\mathbb R$ and $\sigma>0$ are fixed. Positive rescaling and
translation preserve symmetry and unimodality, so
$H_f\in\mathcal M^{us}(\mu,\sigma)$.
\begin{proposition}[Approximation error bounds]\label{pro.5}
Let $L\in L^2(0,1)$, and let $L_n$ be step functions of the form in
Proposition~\ref{pro.4}. Suppose $b_L>0$ and $b_{L_n}>0$, and set
$$
\delta_n=\|{L_n^{\mathrm a}-L^{\mathrm a}}\|.
$$
Then
\begin{align}
&\|{L_{n,us}^{\uparrow}-L_{us}^{\uparrow}}\|
\le\|{L_n-L}\|,\label{eq:projectionerror}\\
&\|{(L_{n,us}^{\uparrow}-a_{L_n})-(L_{us}^{\uparrow}-a_L)}\|
\le\delta_n,\label{eq:centerederror}\\
&\|{h_{L_n}-h_L}\|\le\frac{2\sigma}{b_L}\delta_n.
\label{eq:quantileerror}
\end{align}
If $F$ has finite variance $\sigma_F^2$, then
\begin{equation}\label{eq:budgeterror}
\left|d_W^2(F,H_{L_n})-d_W^2(F,H_L)\right|
\leq\frac{4\sigma\sigma_F}{b_L}\delta_n.
\end{equation}
Moreover, if $g$ is absolutely continuous on $[0,1]$, $g(0)=0$, and
$g'\in L^2(0,1)$, then
\begin{equation}\label{eq:riskerror}
\left|\rho_g(H_{L_n})-\rho_g(H_L)\right|
\le\frac{2\sigma\|{\bar {g}'(1-\cdot)}\|}{b_L}\delta_n
.
\end{equation}
Consequently, if $L_n\to L$ in $L^2(0,1)$ and $b_L>0$, then
$b_{L_n}>0$ for all sufficiently large $n$, and the projected functions,
standardized quantiles and corresponding distortion riskmetrics converge.
\end{proposition}

Propositions~\ref{pro.4} and~\ref{pro.5} adapt the step-function
representation and error estimates of \cite{liu2025robust} to symmetry;
Proposition~\ref{prop:fixed-node} provides a convex approximation on a
fixed grid. With $L=k_\lambda$, Proposition~\ref{pro.5} controls errors
at a fixed multiplier. The next remark translates a bound on the total
projection error into feasibility and optimality guarantees.

\begin{remark}[Feasibility and stopping criteria]\label{rem:stopping}
Suppose $\widetilde p\in\mathcal C\setminus\{0\}$ satisfies
$\notationedit{\|\widetilde p-P_{\mathcal C}k_\lambda^{\mathrm a}\|}\leq r$ and
$\widetilde b=\notationedit{\|\widetilde p\|}>r$. The standardized quantile
$\widetilde h=\mu+\sigma\widetilde p/\widetilde b$ has the prescribed
moments and shape. The preceding error bounds hold with $\delta_n$
replaced by $r$ and $b_L$ by $\widetilde b-r$, provided $r$ includes
the grid, input, solver and integration errors.
For any $\lambda\geq0$, penalization yields
\begin{equation}\label{eq:numerical-upper}
\sup_{G\in\mathcal M^{us}_\epsilon(\mu,\sigma)}\rho_g(G)
\leq U_\lambda:=\mu g(1)+\sigma(\widetilde b+r)-\lambda D_\epsilon,
\qquad
D_\epsilon=\frac{(\mu-\mu_F)^2+\sigma^2+\sigma_F^2-\epsilon}{2}.
\end{equation}
To obtain this bound, use
$\langle F^{-1},h-\mu\rangle\geq D_\epsilon$ for every feasible
quantile and bound the penalized objective by
$\sigma b_{\lambda,us}\leq\sigma(\widetilde b+r)$.
If $\tau_W$ bounds the integration error in the computed squared
distance of $\widetilde h$, feasibility is ensured when that distance
plus $\tau_W$ is at most $\epsilon$. The computed risk, less its
integration error bound $\tau_\rho$, then gives a lower bound $L$.
\notationedit{For a prescribed absolute tolerance $\tau>0$, the condition
$U_\lambda-L\leq\tau$ ensures an optimality gap of at most $\tau$},
even if the multiplier equation is not solved exactly.
For VaR, whose distortion is discontinuous, \eqref{eq:riskerror} does
not apply. Its quantile-envelope formulas are evaluated directly in
Section~\ref{sec5}.
\end{remark}

\section{Numerical examples and simulation}\label{sec5}
This section examines how shape restrictions affect worst-case risk bounds. We first compare bounds for fixed reference distributions and then use simulated samples to calibrate Wasserstein budgets and evaluate the resulting worst-case ES.

To compute the unimodal ES and RVaR bounds, we consider a finite-dimensional cone $\mathcal K_m$ of centered, nondecreasing quantile functions that are concave on $(0,1/2)$ and convex on $(1/2,1)$. For the symmetric unimodal class, we additionally impose antisymmetry about $1/2$. Writing $h=\mu+\sigma y$ and $w(u)=g'(1-u)$, the corresponding risk is $\mu g(1)+\sigma\langle w,y\rangle$. We compute $y$ by solving the second-order cone program
\begin{equation}\label{eq:numerical-socp}
\max_{y\in\mathcal K_m}
\left\{\langle w,y\rangle:
\|y\|\leq1,\quad
\langle F^{-1},y\rangle\geq D_\epsilon/\sigma\right\},
\end{equation}
where $D_\epsilon$ is defined in \eqref{eq:numerical-upper}. Since $y$ is centered and both $y$ and $F^{-1}$ are nondecreasing, $\langle F^{-1},y\rangle\geq0$. Hence, any feasible $y$ with a positive objective value can be rescaled to unit norm without decreasing the objective or violating the covariance constraint, regardless of the sign of $D_\epsilon$. The optimal centered risks are positive in all examples below, so the norm relaxation is exact and the prescribed variance is attained. We solve the programs in Python using CVXPY with the Clarabel solver and check the residuals of the moment, shape, and Wasserstein constraints. 

\subsection{Fixed reference distributions}\label{subsec:fixed-references}
\begin{figure}[htbp!]\centering
\captionsetup{labelfont=bf}
\includegraphics[width=0.9\linewidth]{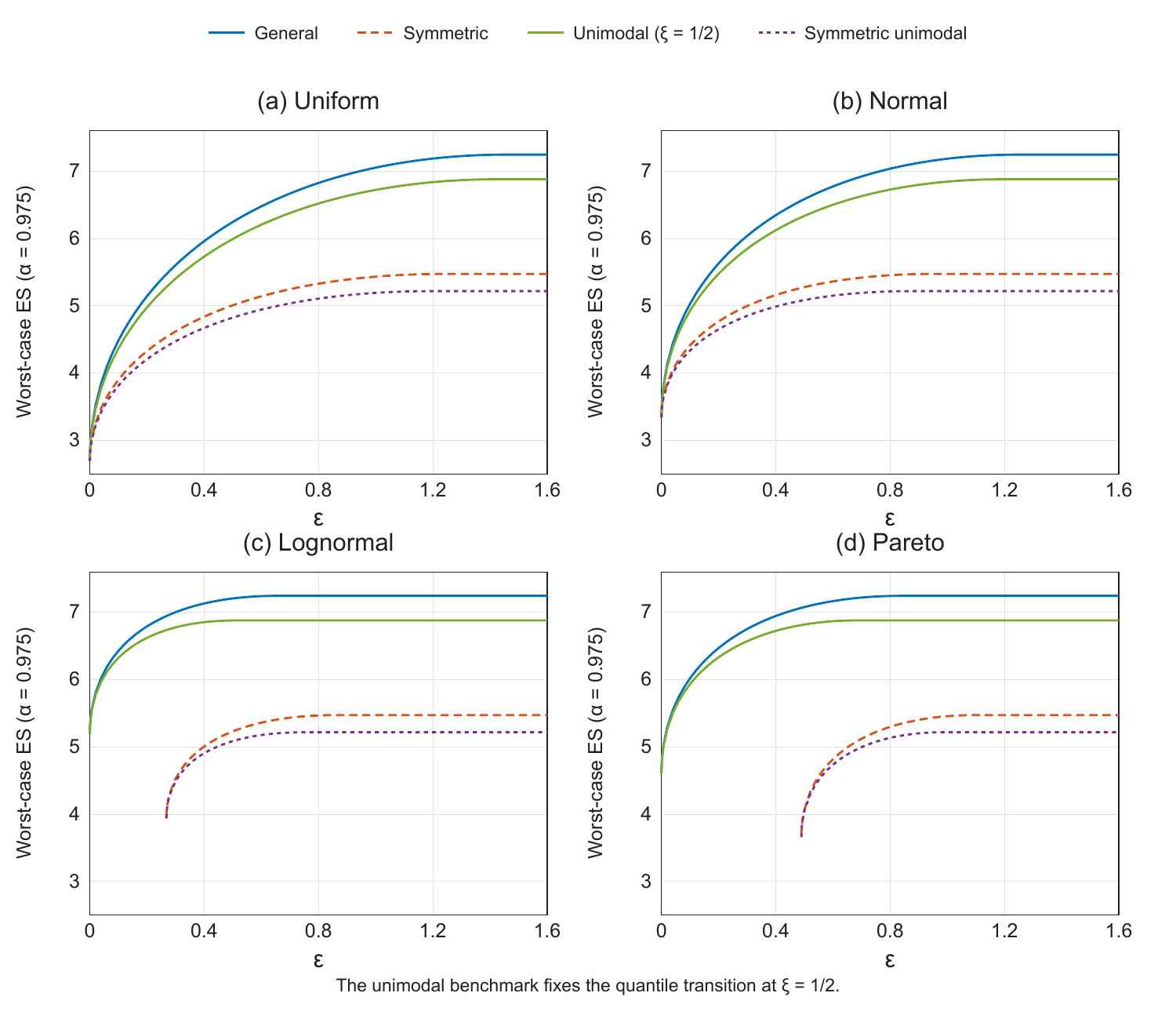}
\caption{\small Worst-case $\ES_{\alpha}$ at $\alpha=0.975$ under
moment and Wasserstein constraints and the indicated shape
restrictions. Curves are shown only for feasible budgets.
The unimodal and symmetric unimodal curves are numerical
approximations; the unimodal benchmark without symmetry fixes
the quantile transition at $\xi=1/2$.}
\label{fig1}
\end{figure}

\begin{figure}[htbp!]\centering
\captionsetup{labelfont=bf}
\includegraphics[width=0.9\linewidth]{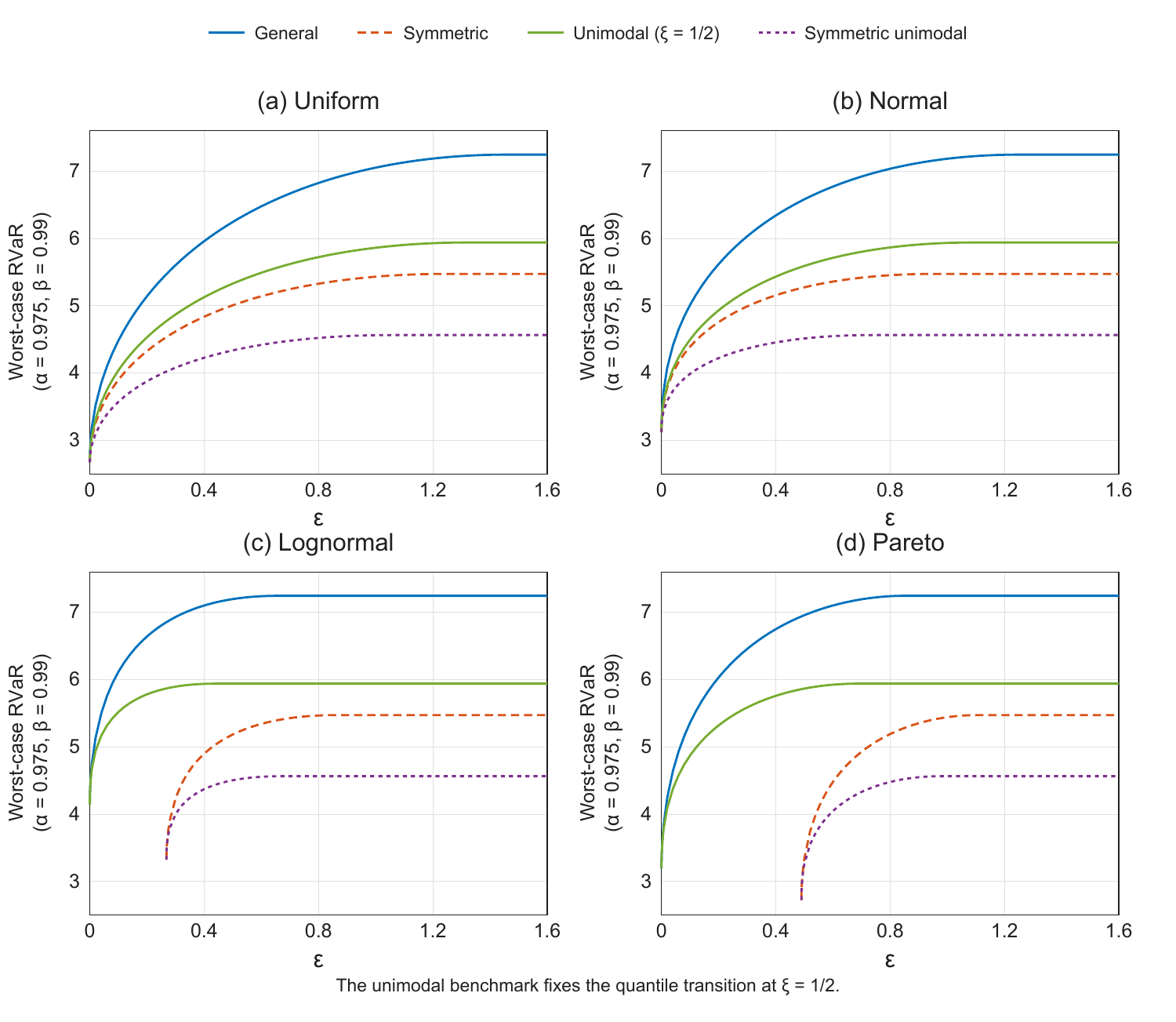}
\caption{\small Worst-case $\RVaR_{(\alpha,\beta)}$ at
$(\alpha,\beta)=(0.975,0.99)$ under the same constraints as in
Fig.~\ref{fig1}. Curves are shown only for feasible budgets.
The unimodal and symmetric unimodal curves are numerical
approximations.}
\label{fig2}
\end{figure}

\begin{figure}[htbp!]\centering
\captionsetup{labelfont=bf}
\includegraphics[width=0.9\linewidth]{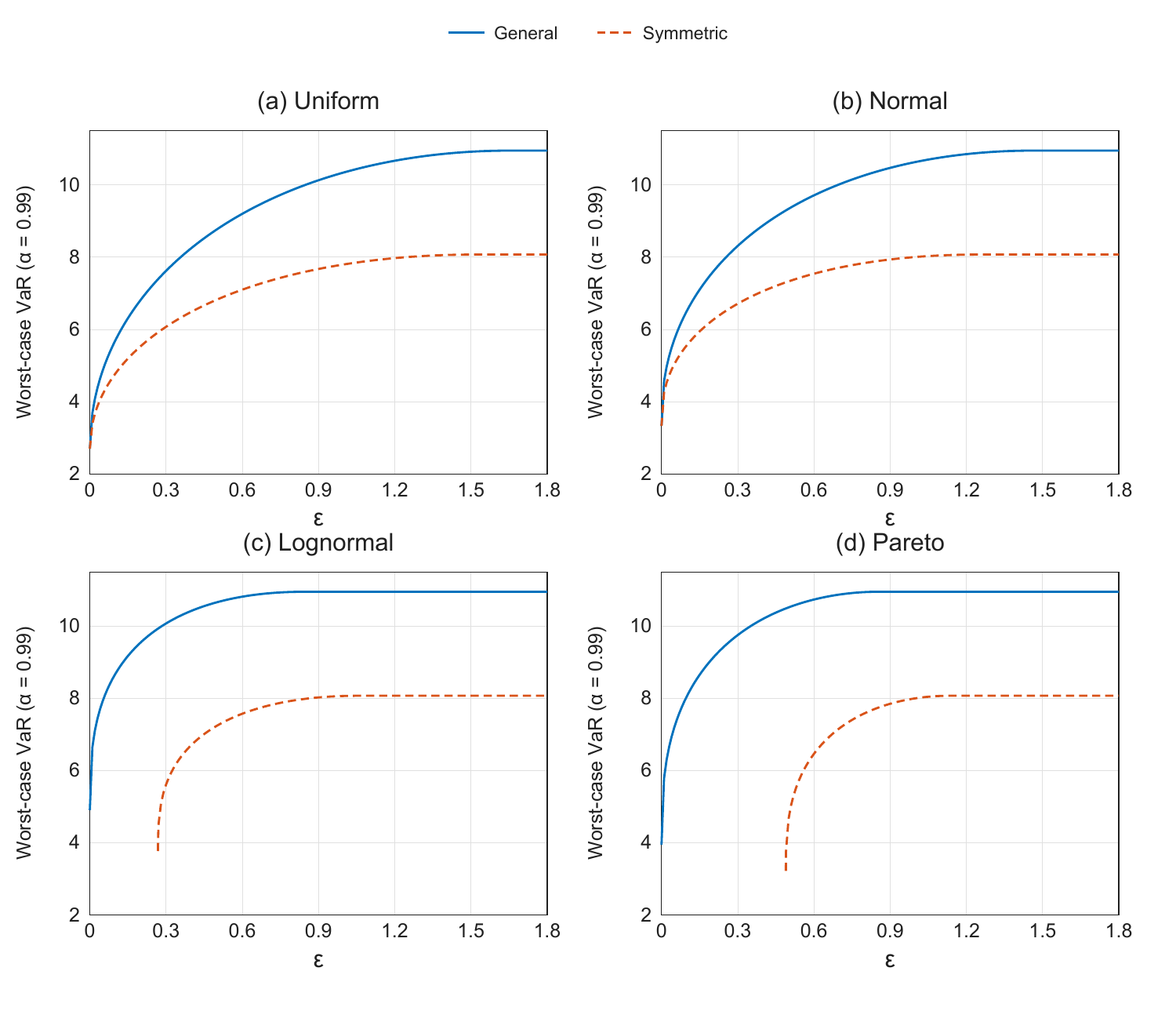}
\caption{\small Worst-case $\VaR_{\alpha}$ at $\alpha=0.99$
with and without symmetry, under moment and Wasserstein constraints.
The curves show suprema; for the symmetric class, these are not
attained above the feasibility threshold (Subsection~\ref{subsec:sym-var}).}
\label{fig3}
\end{figure}

Figures~\ref{fig1}--\ref{fig3} show how the worst-case values of
$\ES_{0.975}$, $\RVaR_{(0.975,0.99)}$, and $\VaR_{0.99}$ vary
with the squared Wasserstein budget $\epsilon$. All reference and
candidate distributions have mean $1$ and standard deviation $1$.
Panels (a)--(d) correspond to the uniform $U(1-\sqrt3,1+\sqrt3)$,
normal $N(1,1)$, lognormal $LN(-(\log2)/2,\sqrt{\log2})$, and
type I Pareto $\operatorname{Pareto}(1+\sqrt2,2-\sqrt2)$ reference
distributions, respectively. The lognormal distribution is
parameterized by the mean and standard deviation of its logarithm,
and the Pareto distribution by its shape and scale. For ES and RVaR,
we compare four classes: general, symmetric, unimodal with the
quantile transition fixed at $\xi=1/2$, and symmetric unimodal.
For VaR, we compare the general and symmetric classes.

The symmetric bounds are obtained from Corollaries~\ref{ex.1},
\ref{ex.5}, and~\ref{ex.6}, while the symmetric unimodal bounds
are computed using Theorem~\ref{th.4} and
Proposition~\ref{prop:fixed-node}. The general ES and RVaR
benchmarks follow \citet{bernard2024robust}, and the unimodal
benchmarks with a fixed quantile transition follow
\citet{liu2025robust}. We evaluate the general and symmetric
bounds using envelope formulas and isotonic pooling, with
one-dimensional root searches when the Wasserstein constraint
is active. For the unimodal calculations in
\eqref{eq:numerical-socp}, we use piecewise-linear quantiles;
in the symmetric unimodal case, the approximation cones are
those defined in Proposition~\ref{prop:fixed-node}.

In these examples, symmetry yields tighter ES and RVaR bounds
than unimodality with the transition fixed at $\xi=1/2$,
when compared at common feasible budgets. This observation
is specific to the examples considered here; the two shape
classes are not nested. The symmetric unimodal class is
contained in both, so its sharp bound is no greater than
that of either class whenever the joint constraints are feasible.

The reference distribution affects both the minimum budget
required for feasibility and the budget at which the bound
reaches the plateau determined solely by the moment and shape
constraints. In particular, an asymmetric reference entails
a strictly positive minimum budget when symmetry is imposed.
For the lognormal reference in Fig.~\ref{fig1}, the symmetric
ES curve begins at $\epsilon_0=2-\sqrt3\approx0.2679$ and
reaches its plateau at $\epsilon_1\approx0.8626$, whereas the
general class is feasible from $\epsilon=0$. These two
thresholds illustrate, respectively, the feasibility
characterization in Lemma~\ref{lem:sym-feasibility} and
the saturation regime in Theorem~\ref{th.3}.

\subsection{ES-based budget calibration and shape restrictions}\label{subsec:calibration}

Let $X_1,\ldots,X_n$ be independent observations from a sampling
distribution $F^\star$, with empirical distribution
$F_n=n^{-1}\sum_{i=1}^n\delta_{X_i}$. For a reference distribution
$F$, let $B_j(\epsilon;F)$ denote the worst-case $\ES_{0.975}$
under the prescribed moment constraints and squared Wasserstein
budget $\epsilon$, where $j\in\{\mathrm{gen},s,us\}$ indexes the
general, symmetric, and symmetric unimodal classes, respectively.
We set $B_j(\epsilon;F)=-\infty$ whenever the corresponding
class is empty.

Data-driven calibration of robustness parameters has been
studied using bootstrap methods \citep{GotohKimLim2021}
and cross-validation \citep{HanasusantoKuhn2018}.
Drawing on the bootstrap validation approach of
\citet[Section~7.2.3]{MohajerinEsfahaniKuhn2018}
and the related moment-based procedure of
\citet[Section~4.3]{TangYang2023}, we calibrate the Wasserstein
budget by comparing training-sample worst-case ES bounds
with ES estimates from the corresponding out-of-bag
validation samples.

For each original sample, we draw $B=500$ bootstrap resamples
of size $n$. For resample $b$, let $F_n^{*(b)}$ denote the
training empirical distribution and $F_{\mathrm{val}}^{(b)}$
the empirical distribution of the out-of-bag observations,
namely those whose indices are absent from the resample.
The validation ES estimate is
\[
V_b=\frac{1}{1-\alpha}\int_\alpha^1
\bigl(F_{\mathrm{val}}^{(b)}\bigr)^{-1}(u)\,du,
\qquad \alpha=0.975.
\]
We account for fractional weight at the tail boundary when
evaluating this empirical quantile integral. Holding the
bootstrap resamples fixed throughout calibration, we define
the empirical validation rate and the selected budget by
\begin{equation}\label{eq:dd-coverage-target}
\widehat C_{\mathrm{gen}}(\epsilon)
=\frac1B\sum_{b=1}^B
\mathbf1\{B_{\mathrm{gen}}(\epsilon;F_n^{*(b)})\ge V_b\},
\qquad
\widehat\epsilon_n
=\inf\{\epsilon\ge0:\widehat C_{\mathrm{gen}}(\epsilon)\ge0.95\}.
\end{equation}
The convention for empty classes ensures that infeasible
training problems count as validation failures. Since
$B_{\mathrm{gen}}(\epsilon;F)$ is nondecreasing in $\epsilon$
for each fixed reference $F$, the calibration rule can
equivalently be expressed as
\begin{equation}\label{eq:dd-bootstrap}
\epsilon_b^{\mathrm{req}}
=\inf\{\epsilon\ge0:
B_{\mathrm{gen}}(\epsilon;F_n^{*(b)})\ge V_b\},
\qquad
\widehat\epsilon_n=\epsilon_{(475)}^{\mathrm{req}}.
\end{equation}
Here, $\epsilon_{(475)}^{\mathrm{req}}$ denotes the 475th
smallest of the $B=500$ required budgets, with the convention
$\inf\varnothing=+\infty$. For each resample, we compute
$\epsilon_b^{\mathrm{req}}$ by scalar inversion of the general
ES formula used in Subsection~\ref{subsec:fixed-references},
starting at the corresponding feasibility threshold.
If $V_b$ exceeds the moment-only ES bound
$\mu+\sigma\sqrt{\alpha/(1-\alpha)}$, then
$\epsilon_b^{\mathrm{req}}=+\infty$. If
$\epsilon_{(475)}^{\mathrm{req}}=+\infty$, no finite budget
satisfies the calibration criterion, and calibration is
declared unsuccessful. Otherwise, the selected Wasserstein
radius is $\sqrt{\widehat\epsilon_n}$.

Following successful calibration, we evaluate all three classes
using the full-sample reference $F_n$ and the common budget
$\widehat\epsilon_n$. We record infeasibility at this budget
rather than enlarging it, so that the comparison isolates the
effect of the shape restrictions. The 95\% criterion in
\eqref{eq:dd-coverage-target} is a bootstrap validation target.
It does not, by itself, imply a 95\% probability that the
calibrated ball contains $F^\star$ or that a computed bound
covers its ES. Nor does calibration based on the general class
ensure 95\% validation coverage for either shape-restricted
class. Population ES coverage is assessed separately through
the simulation repetitions described below.

We consider five sampling distributions with mean zero and
variance one:
\[
N(0,1),\qquad \sqrt{3/5}\,T_5,\qquad
\operatorname{Laplace}(0,1/\sqrt2),\qquad
\frac{Z+2S}{\sqrt5},\qquad
\frac{e^Z-e^{1/2}}{\sqrt{e^2-e}},
\]
where $T_5$ follows a Student-$t$ distribution with five degrees
of freedom, $Z\sim N(0,1)$, and $S$ is independent of $Z$ and
takes values $\pm1$ with equal probability. The Laplace
parameters denote location and scale. The first three
distributions are symmetric and unimodal. The fourth is
symmetric but bimodal, and the fifth is asymmetric. For each
distribution and each sample size $n\in\{250,500,1000,2000\}$,
we generate 1000 independent samples and repeat the full
calibration and evaluation procedure for each sample.
The population moments $(\mu,\sigma)=(0,1)$ are treated as
known throughout; neither the training nor the validation
samples are restandardized.

We compute the general and symmetric ES bounds using the
formulas in Subsection~\ref{subsec:fixed-references}. For
symmetric unimodal ES, we solve \eqref{eq:numerical-socp}
over continuous piecewise-linear functions that vanish at
$1/2$, are nondecreasing and convex on $[1/2,1]$, and are
extended antisymmetrically to $[0,1/2]$. The grid on
$[1/2,1]$ comprises 40 equally spaced intervals on each of
$[0.5,0.9]$ and $[0.9,0.975]$, 48 geometrically spaced
intervals from $0.975$ to $1-10^{-5}$, and additional nodes
at $1$ and $0.9625$. The node at $0.9625$ is the hinge of
the symmetric unimodal ES optimizer under the moment
constraints alone at $\alpha=0.975$. The resulting grid
has 131 nodes, with no additional tail function. The
integrals defining the norm and ES objective are evaluated
exactly within this family. Covariance integrals are
likewise exact after splitting at the jumps of the
empirical quantile.

Maximizing covariance with $F_n^{-1}$ under the same shape
and norm constraints identifies the nearest quantile in
this finite-dimensional family. The corresponding minimum
squared distance provides an upper bound on the exact
feasibility threshold for the full symmetric unimodal
class. Accordingly, the symmetric unimodal feasibility
results reported below refer to the finite-dimensional
approximation. We assess sensitivity to the grid by
doubling its resolution for the first ten samples of
each distribution and sample-size combination, giving
200 comparisons in total. Feasibility classifications
agree across all 200 comparisons. Among the
$\simGridFeasiblePairs$ pairs that are feasible on both
grids, the largest change in ES is $\simGridChange$.

We report calibration success, feasibility, and population
ES coverage across all 1000 repetitions. Calibration
failures and unavailable bounds are counted as noncoverage.
Mean risk estimates are computed on the common subset of
repetitions for which all three calculations are feasible.
The coverage curves include pointwise 95\% Wilson intervals
to quantify Monte Carlo uncertainty. The out-of-bag
validation sample contains approximately $n/e$ observations
on average. At $n=250$, its upper 2.5\% tail therefore
corresponds to only about 2.3 observations on average.
The smallest-sample setting thus also tests the stability
of calibration based on a small number of tail observations.
The outer repetitions assess the resulting population ES
coverage.

\begin{table}[htbp!]\centering
\small
\setlength{\tabcolsep}{3.5pt}
\ifshowrevisions\color{studybrown}\captionsetup{labelfont={bf,color=studybrown},textfont={color=studybrown}}\fi
\caption{\small Calibration, feasibility, and population ES
coverage at $n=1000$. Median squared budgets are conditional
on successful calibration; all other entries are percentages
of the 1000 repetitions. Feasibility for $us$ refers to the
finite-dimensional approximation. Calibration failures and
unavailable bounds are counted as noncoverage.}
\label{tab:simulation-coverage}
\begin{tabular}{@{}lrrrrrrrr@{}}
\toprule
& Median & Calibration & \multicolumn{3}{c}{Feasible} & \multicolumn{3}{c}{ES coverage}\\
\cmidrule(lr){4-6}\cmidrule(l){7-9}
Distribution & $\widehat\epsilon_n$ & success & $\mathrm{gen}$ & $s$ & $us$ & $\mathrm{gen}$ & $s$ & $us$\\
\midrule
Normal & 0.0083 & 100.0 & 100.0 & 100.0 & 100.0 & 100.0 & 100.0 & 100.0\\
Student-$t_5$ & 0.0194 & 100.0 & 100.0 & 99.0 & 98.6 & 100.0 & 98.9 & 98.4\\
Laplace & 0.0173 & 100.0 & 100.0 & 99.4 & 99.4 & 100.0 & 99.4 & 99.4\\
Bimodal & 0.0060 & 100.0 & 100.0 & 100.0 & 1.6 & 100.0 & 100.0 & 1.6\\
Lognormal & 0.0884 & 78.7 & 78.7 & 1.2 & 1.1 & 78.7 & 0.0 & 0.0\\
\bottomrule
\end{tabular}

\end{table}

For the three correctly specified distributions, the
calibrated budgets and mean risk bounds decrease with
sample size. For the normal distribution, the median
squared budget falls from $\simNormalBudgetSmall$ at
$n=250$ to $\simNormalBudgetLarge$ at $n=2000$. At
$n=1000$, imposing symmetry reduces mean worst-case ES
by $\simNormalSymmetricReduction$--$\simStudentSymmetricReduction\%$
relative to the general class, using the common feasible
repetitions. Adding unimodality yields a further
$\simNormalUnimodalReduction$--$\simStudentUnimodalReduction\%$
reduction relative to the symmetric class. Most of the
observed tightening therefore comes from symmetry.

\begin{figure}[htbp!]\centering
\includegraphics[width=0.9\linewidth]{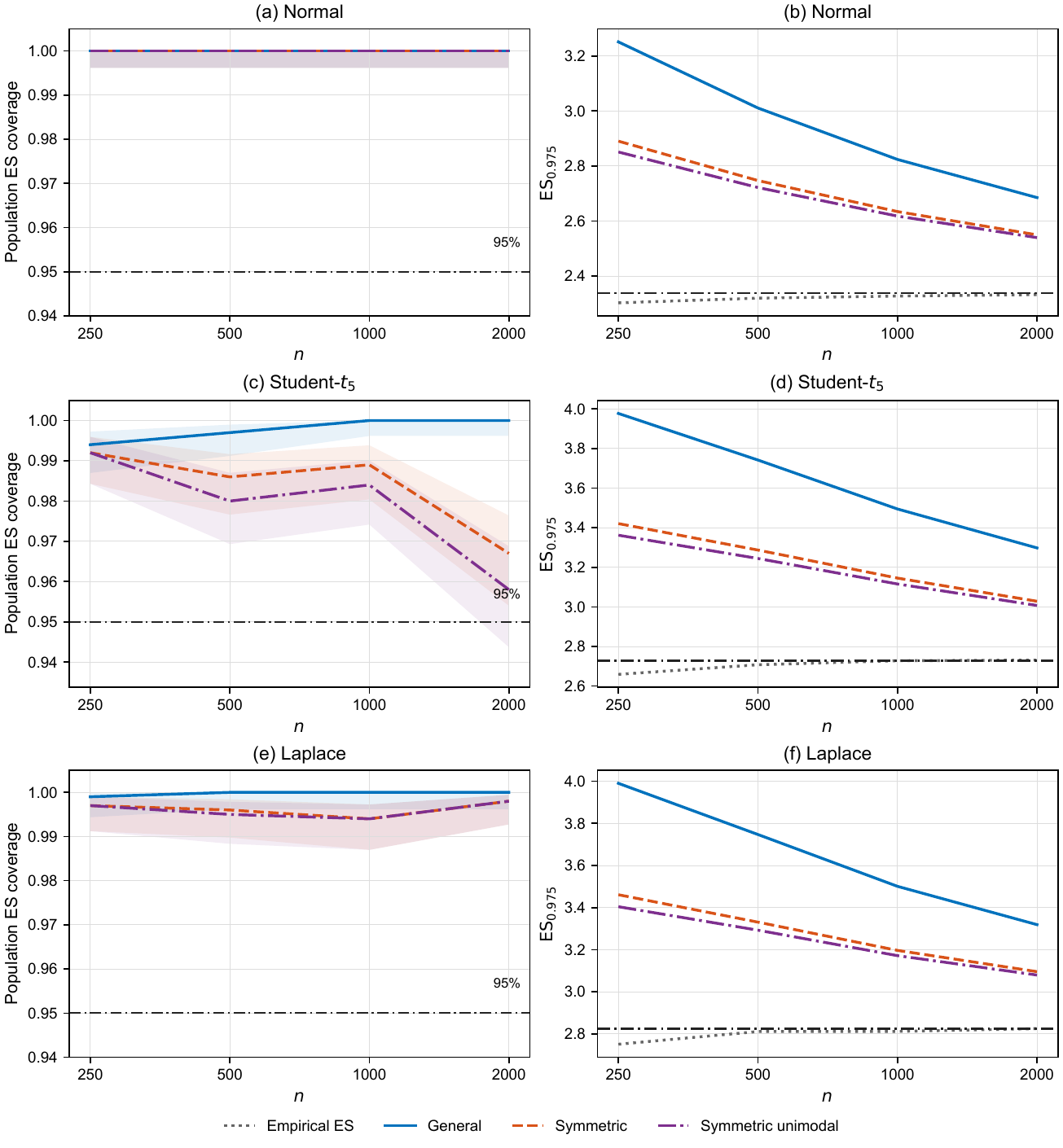}
\ifshowrevisions\captionsetup{labelfont={bf,color=studybrown},textfont={color=studybrown}}\fi
\caption{\small ES estimates under a common calibrated
Wasserstein budget. Left: population ES coverage across
1000 independent samples, with pointwise 95\% Wilson
intervals and a 95\% reference line. Right: mean empirical
and worst-case ES over repetitions for which all three
calculations are feasible; the horizontal line indicates
the population ES. Symmetric unimodal values are computed
over the finite-dimensional feasible family. The $n$-axis
is logarithmic.}
\label{fig:simulation}
\end{figure}

Calibration based on the general class produces conservative
bounds in these experiments: observed population ES coverage
is 100\% for all three correctly specified distributions at
$n=1000$ (Table~\ref{tab:simulation-coverage}). Shape
restrictions yield smaller bounds while retaining high
observed coverage. For the Student-$t_5$ distribution at
$n=2000$, coverage under symmetry and under symmetric
unimodality is $\simStudentSymmetricCoverageLarge\%$ and
$\simStudentJointCoverageLarge\%$, respectively
(Figure~\ref{fig:simulation}). These results illustrate
the reduction in observed coverage that can accompany
tighter bounds under a common calibrated budget.

The two misspecified cases reveal distinct limitations.
At $n=1000$, the symmetric calculation is feasible in
$\simBimodalSymmetricFeasible\%$ of repetitions for the
bimodal distribution, whereas the symmetric unimodal
calculation is feasible in only
$\simBimodalJointFeasible\%$. For the lognormal distribution,
calibration succeeds in $\simLognormalCalibrationSuccess\%$
of repetitions. Validation ES estimates can exceed the
moment-only cap because the empirical validation
distributions need not satisfy the prescribed population
moments. If such exceedances occur in more than 5\% of
the bootstrap resamples, no finite budget can meet the
calibration target. With calibration failures counted
as noncoverage, population ES coverage for the general
class is $\simLognormalGeneralCoverage\%$. At this sample
size, imposing the incorrect symmetry restriction yields
zero observed coverage.

\section{Conclusion}\label{sec:conclusion}
We show how symmetry affects both the minimum feasible
Wasserstein budget and the worst-case risk under prescribed
mean and variance. Under the stated regularity and
nondegeneracy assumptions, we characterize sharp upper
bounds for distortion functions of bounded variation
through the concave envelope of a symmetrized auxiliary
distortion. Our approximation argument preserves symmetry
and the prescribed first two moments exactly, establishing
that upper semicontinuous relaxation of the symmetrized
distortion leaves the supremum unchanged for budgets
strictly above the feasibility threshold, even when the
supremum is not attained. With unimodality imposed in
addition to symmetry, we obtain an $L^2$-projection
characterization for absolutely continuous distortions
with square-integrable derivatives. Finite-dimensional
representations and error estimates provide a basis
for numerical approximation.

The numerical examples illustrate how the reference
distribution determines the feasibility threshold and
affects the worst-case bound while the Wasserstein
constraint is active. Beyond the saturation threshold,
the bound depends only on the prescribed moments and
shape restrictions. \studyedit{The simulations compare
the shape classes at a common Wasserstein budget
calibrated through ES-based validation and assess
population ES coverage across independent repetitions.
They quantify the tightening achieved with correctly
specified shape restrictions and illustrate how
misspecification can lead to infeasibility or lower
coverage.}

\section*{Acknowledgements}
\begingroup\emergencystretch=3em
Yiying Zhang acknowledges financial support from the National Natural
Science Foundation of China (No. 12571513) and the Shenzhen Science and
Technology Program (grant
JCYJ20250604144210013).  GPT-6.1 Sol was used to assist with coding the numerical experiments and language writing polishing.
\par\endgroup

\bibliographystyle{mystyle-main7}
\setlength{\bibsep}{0.0pt}
\bibliography{literature}

\appendix

\section{Proofs of the main results in Section~\ref{sec3}}\label{app:symmetry}
\subsection{Proof of Proposition~\ref{prop.1}}
\begin{proof}
We first write \eqref{b2} in the equivalent form
\begin{align}\label{b1}
 \rho_{g}(X)=\int_{-\infty}^{0}\left[g\left(\mathbb{P}(X> x)\right)-g(1)\right]\mathrm{d}x+\int_{0}^{\infty}g\left(\mathbb{P}(X> x)\right)\mathrm{d}x,
\end{align}
where $X\sim G$.
Proposition~2(iii) of \cite{Wang2020distortion} gives
\begin{align*}
\rho_{g}(X)&=\mu g(1)+\rho_{g}(X-\mu).
\end{align*}
It therefore remains to show that $\rho_{g}(X-\mu)=\rho_{\bar{g}}(G)$.
By definition and a change of variables,
\begin{align*}
\rho_{g}(X-\mu)&=\int_{-\infty}^{0}\left[g\left(\mathbb{P}(X-\mu> x)\right)-g(1)\right]\mathrm{d}x+\int_{0}^{\infty}g\left(\mathbb{P}(X-\mu> x)\right)\mathrm{d}x\\
&=\int_{0}^{\infty}\left[g\left(\mathbb{P}(X-\mu> -x)\right)-g(1)\right]\mathrm{d}x+\int_{-\infty}^{0}g\left(\mathbb{P}(X-\mu> -x)\right)\mathrm{d}x.
\end{align*}
Averaging these two representations yields
\begin{align}\label{b3}
\nonumber\rho_{g}(X-\mu)&=\frac{1}{2}\bigg\{\int_{-\infty}^{0}\left[g\left(\mathbb{P}(X-\mu> x)\right)-g(1)\right]\mathrm{d}x+\int_{0}^{\infty}g\left(\mathbb{P}(X-\mu> x)\right)\mathrm{d}x\\
\nonumber&~~~~+\int_{0}^{\infty}\left[g\left(\mathbb{P}(X-\mu> -x)\right)-g(1)\right]\mathrm{d}x+\int_{-\infty}^{0}g\left(\mathbb{P}(X-\mu> -x)\right)\mathrm{d}x\bigg\}\\
&=\int_{-\infty}^{\infty}\frac{g(\mathbb{P}(X-\mu>x))+g(\mathbb{P}(X-\mu>-x))-g(1)}{2}\mathrm{d}x.
\end{align}
Since $X$ is symmetric about $\mu$, we have $\mathbb{P}(X-\mu>x)=\mathbb{P}(X-\mu<-x)$.
Thus \eqref{b3} becomes
\begin{align}\label{b4}
\nonumber\rho_{g}(X-\mu)&=\int_{-\infty}^{\infty}\frac{g(\mathbb{P}(X-\mu<-x))+g(\mathbb{P}(X-\mu>-x))-g(1)}{2}\mathrm{d}x\\
&=\int_{-\infty}^{\infty}\frac{g(\mathbb{P}(X<x))+g(\mathbb{P}(X>x))-g(1)}{2}\mathrm{d}x.
\end{align}
On the other hand,
\begin{align}\label{b5}
\nonumber\rho_{\bar{g}}(G)&=\int_{-\infty}^{\infty}\frac{g(1-G(x))+g(G(x))-g(1)}{2}\mathrm{d}x\\
&=\int_{-\infty}^{\infty}\frac{g(\mathbb{P}(X>x))+g(\mathbb{P}(X\leq x))-g(1)}{2}\mathrm{d}x.
\end{align}
The integrands in \eqref{b4} and \eqref{b5} agree except possibly at the at most countably many atoms of $G$. These points do not affect the Lebesgue integrals. Comparing \eqref{b4} and \eqref{b5} gives $\rho_{g}(X-\mu)=\rho_{\bar{g}}(G)$. Hence $\rho_{g}(G)=\mu g(1)+\rho_{\bar{g}}(G)$, as required.\qedhere
\end{proof}

\subsection{Proof of Lemma~\ref{lem:sym-feasibility}}
\begin{proof}
For $G\in\mathcal M^s(\mu,\sigma)$, set $x=G^{-1}-\mu$.
Symmetry gives $x(1-u)=-x(u)$ almost everywhere and hence
\[
\langle F^{-1},x\rangle=\langle s_F,x\rangle,\qquad
 d_W^2(F,G)=A_F-2\langle s_F,x\rangle
 \geq A_F-2\sigma\sigma_0.
\]
Since $F^{-1}$ is nonconstant and nondecreasing, $\sigma_0>0$.
Equality in Cauchy--Schwarz holds precisely when
$x=\sigma s_F/\sigma_0$ almost everywhere, which proves the empty
and singleton cases. This function is a nondecreasing, antisymmetric
quantile, so the lower bound is attained at $H_\infty$.
For $\epsilon>\epsilon_0$, take a different centered symmetric
quantile $z$ with $\notationedit{\|z\|}=\sigma$, and normalize
$(1-t)(H_\infty^{-1}-\mu)+tz$ to norm $\sigma$.
For sufficiently small $t>0$, the resulting quantile differs from
$H_\infty^{-1}$ and has squared distance below $\epsilon$.
\end{proof}

\subsection{Proof of Lemma~\ref{lem:sym-envelope-regularity}}
\begin{proof}
Endpoint continuity of $g$ implies endpoint continuity of $g^*$
and $\bar g^*$; see \citet[Section~2.2]{Pesenti2025robust}.
The reference term in $g_\lambda$ is absolutely continuous,
so $g_\lambda$ and its concave envelope are also continuous at
both endpoints. Since $\bar g(0)=\bar g(1)=g_\lambda(0)=g_\lambda(1)=0$,
their concave envelopes vanish at both endpoints. Each envelope
is therefore a finite concave function continuous on $[0,1]$,
and hence is absolutely continuous there.

We use the following derivative comparison.
 If $c$ and $d$ are
absolutely continuous concave functions vanishing at both endpoints,
with $c\leq d$, then, for $\alpha\in(0,1)$,
\[
\int_\alpha^1c'(1-u)\,\mathrm du
=c(1-\alpha)
\leq d(1-\alpha)
=\int_\alpha^1d'(1-u)\,\mathrm du.
\]
Both derivatives have integral zero, and $c'(1-u)$ and $d'(1-u)$
are nondecreasing in $u$.
\notationedit{Here $X\leq_{cx}Y$ denotes convex order:
$\mathbb E[\phi(X)]\leq\mathbb E[\phi(Y)]$ for all convex $\phi$
with well-defined expectations.}
The quantile characterization of convex order gives
$c'(1-U)\leq_{cx}d'(1-U)$. Consequently,
\begin{equation}\label{eq:concave-derivative-comparison}
\notationedit{\|c'\|}\leq\notationedit{\|d'\|}
\end{equation}
whenever $d'\in L^2(0,1)$.

Define
\[
B(t)=\frac{g^*(t)+g^*(1-t)-g(1)}2.
\]
The function $B$ is concave, vanishes at both endpoints, and
satisfies $\bar g\leq\bar g^*\leq B$. Its derivative is
$B'(t)=\bigl((g^*)'(t)-(g^*)'(1-t)\bigr)/2$ almost everywhere.
Applying \eqref{eq:concave-derivative-comparison} and the triangle
inequality yields
\begin{equation}\label{eq:bar-envelope-l2}
\notationedit{\|(\bar g^*)'\|}
\leq\notationedit{\|B'\|}
\leq\notationedit{\|(g^*)'\|}<\infty.
\end{equation}

Next, set
\[
R_F(t)=\int_{1-t}^1F^{-1}(s)\,\mathrm ds
+\int_t^1F^{-1}(s)\,\mathrm ds-\mu_F.
\]
This function is concave and absolutely continuous, with zero
endpoint values and
\[
{R_F'(t)=F^{-1}(1-t)-F^{-1+}(t)\in L^2(0,1).}
\]
Since $\bar g^*+\lambda R_F$ is a concave majorant of
$g_\lambda=\bar g+\lambda R_F$, we have
$g_\lambda^*\leq\bar g^*+\lambda R_F$.
Another application of \eqref{eq:concave-derivative-comparison}
gives, for every $\lambda\geq0$,
\[
\notationedit{\|(g_\lambda^*)'\|}
\leq\notationedit{\|(\bar g^*)'+\lambda R_F'\|}
\leq\notationedit{\|(\bar g^*)'\|}+\lambda\notationedit{\|R_F'\|}<\infty.
\]
Finally, absolute continuity and the zero endpoint values give
\eqref{eq:sym-envelope-means} by the fundamental theorem of calculus.
\end{proof}

\subsection{Proof of Lemma~\ref{lem:sym-usc-approximation}}
\begin{proof}
Write $k=\bar g$ and
\[
 D(t)=\frac{g_*(t)+g_*(1-t)-g(1)}2,\qquad
 B(t)=\frac{g^*(t)+g^*(1-t)-g(1)}2.
\]
Then $D\leq k\leq\hat k\leq B$, where $D$ is convex, $B$ is
concave, both vanish at the endpoints, and $D',B'\in L^2(0,1)$.
In particular, the survival-integral representations of
$\rho_k$ and $\rho_{\hat k}$ are absolutely integrable on
$\mathcal M^2$. These bounds also show that $k,\hat k\in\mathcal G_2$.

First suppose $X\sim G$ is supported on $[\mu-M,\mu+M]$.
Let $r(x)=\mathbb{P}(X>\mu+x)$ for $x\geq0$, so $0\leq r\leq1/2$.
Symmetry and $k(1-t)=k(t)$ give
\[
\rho_k(G)=2\int_0^M k(r(x))\,\mathrm dx,
\qquad
\rho_{\hat k}(G)=2\int_0^M\hat k(r(x))\,\mathrm dx.
\]
\notationedit{Let $\mathrm{id}(p)=p$ be the identity map on $[0,1/2]$,
and write $\|f\|_\infty:=\sup_{p\in[0,1/2]}|f(p)|$ for the uniform norm.}
We construct continuous, strictly increasing maps
$T_n:[0,1/2]\to[0,1/2]$ satisfying
\[
T_n(0)=0,\qquad
\|T_n-\mathrm{id}\|_\infty\longrightarrow0,\qquad
k(T_n(p))\longrightarrow\hat k(p)
\quad(p\in[0,1/2]).
\]
Because $k$ has bounded variation, its discontinuities in $(0,1/2)$
form a countable set. Let $E_n$ consist of the first $n$ such points
(or all of them if there are fewer) together with $1/2$.
Around the finitely many points of $E_n$, choose pairwise disjoint
neighborhoods, of radius less than $1/n$, whose closures do not
contain $0$ and occur in the same order as the points.
Use a left-sided neighborhood at $1/2$.
For every $p\in E_n$, select an image $y_{n,p}$ in its neighborhood
such that
\[
|y_{n,p}-p|<1/n,\qquad
|k(y_{n,p})-\hat k(p)|<1/n.
\]
This is possible by choosing $p$ itself or approaching it from a side
whose limit realizes $\hat k(p)$. At $p=1/2$, symmetry gives
$k(1/2-)=k(1/2+)$, so $y_{n,1/2}\leq1/2$ suffices; the endpoint
image is allowed to be strictly smaller than $1/2$.
The disjoint neighborhoods ensure that the selected images are
strictly ordered. Linear interpolation through $(0,0)$ and the
points $(p,y_{n,p})$ defines $T_n$. It is strictly increasing and
satisfies $\|T_n-\mathrm{id}\|_\infty<1/n$.
Every discontinuity is eventually selected, whereas convergence
at every continuity point follows from $T_n(p)\to p$.
At $p=0$, use $k(0)=\hat k(0)=0$.

Define $X_n$ by assigning its mass above $\mu$ the survival function
$\mathbb{P}(X_n>\mu+x)=T_n(r(x))$ for $x\geq0$, reflecting that mass
about $\mu$, and placing the remaining mass at $\mu$.
The transformed survival function is right-continuous and
nonincreasing, and its value at zero is at most $1/2$, so this
defines a symmetric law supported on $[\mu-M,\mu+M]$.
Uniform convergence of $T_n$ to the identity gives weak convergence
to $G$; the common bounded support also gives convergence in $d_W$
and $\sigma_n^2:=\operatorname{Var}(X_n)\to\sigma^2$.
Dominated convergence in the preceding survival integrals yields
$\rho_k(X_n)\to\rho_{\hat k}(G)$.
For all sufficiently large $n$, set
\[
Y_n=\mu+\frac{\sigma}{\sigma_n}(X_n-\mu).
\]
Its law has the required symmetry and exact moments.
\notationedit{Writing $\mathcal L(X)$ for the distribution function
$x\mapsto\mathbb P(X\leq x)$ of a random variable $X$, we have}
$d_W(\mathcal L(Y_n),\mathcal L(X_n))\leq|\sigma-\sigma_n|$, and
positive homogeneity together with $k(1)=0$ gives
$\rho_k(Y_n)=(\sigma/\sigma_n)\rho_k(X_n)$.
Thus both required limits hold for $\mathcal L(Y_n)$.

For unbounded $X$, let
\[
X^{(M)}=\mu+\operatorname{sgn}(X-\mu)
              \min\{|X-\mu|,M\},\qquad
\tau_M^2=\operatorname{Var}(X^{(M)}).
\]
Symmetric clipping gives
$\notationedit{\|X^{(M)}-X\|}\to0$ and $\tau_M\to\sigma>0$.
Its survival function agrees with that of $X$ for $0\leq x<M$, so
absolute integrability gives
\[
\rho_{\hat k}(X^{(M)})
=2\int_0^M\hat k(r(x))\,\mathrm dx
\longrightarrow\rho_{\hat k}(G).
\]
For large $M$, normalize by
$Z_M=\mu+\sigma(X^{(M)}-\mu)/\tau_M$.
Then $\mathcal L(Z_M)\in\mathcal M^s(\mu,\sigma)$,
$d_W(\mathcal L(Z_M),G)\to0$, and
$\rho_{\hat k}(Z_M)=(\sigma/\tau_M)\rho_{\hat k}(X^{(M)})
\to\rho_{\hat k}(G)$.
Apply the bounded construction to each $Z_M$ and choose a diagonal
sequence whose Wasserstein and risk approximation errors are both
at most $1/M$. This proves the result.
\end{proof}

\subsection{Proof of Lemma~\ref{le.5}}
\begin{proof}
Proposition~\ref{prop.1} and the envelope bound give
$\rho_g(G)\leq\mu g(1)+\sigma b_0$ on
$\mathcal M^s(\mu,\sigma)$. If $b_0>0$, the moment-constrained
concave-envelope result applied to $\hat{\bar g}$ gives
$\rho_{\hat{\bar g}}(H_0)=\sigma b_0$; see
\citet[Theorem~5 and Lemma~A.2]{Pesenti2025robust}.
Its equality condition specifies $H_0$ uniquely, and the symmetry
of $(\bar g^*)'$ places $H_0$ in $\mathcal M^s(\mu,\sigma)$.
Lemma~\ref{lem:sym-usc-approximation} shows that $\rho_{\bar g}$
has the same supremum.
 If $\bar g=\hat{\bar g}$, it is attained
uniquely at $H_0$.

If $b_0=0$, then $\bar g\leq0$. With $D$ as in the preceding proof,
$D(t)\leq\bar g(t)\leq0$ and
$|D(t)|\leq\sqrt t\,\notationedit{\|D'\mathbf1_{(0,t)}\|}=o(\sqrt t)$.
For $n\geq2$, let
\begin{equation}\label{eq:zero-envelope-sequence}
P_n=\left(1-\frac1n\right)\delta_\mu
 +\frac1{2n}\delta_{\mu-\sigma\sqrt n}
 +\frac1{2n}\delta_{\mu+\sigma\sqrt n}.
\end{equation}
Then $P_n\in\mathcal M^s(\mu,\sigma)$ and
$\rho_{\bar g}(P_n)=2\sigma\sqrt n\,\bar g(1/(2n))\to0$,
which proves sharpness. Finally, $\bar g\equiv0$ makes the objective
constant by Proposition~\ref{prop.1}.
\end{proof}

\subsection{Proof of Lemma~\ref{le.4}}
\begin{proof}
The function $g_\lambda$ is symmetric, upper semicontinuous, and
vanishes at both endpoints. Its concave envelope has derivative
in $L^2$ by Lemma~\ref{lem:sym-envelope-regularity}; its convex
minorant is bounded below by the function $D$ used above, since
$R_F\geq0$. Thus $g_\lambda\in\mathcal G_2$.
Apply the nonconstant-envelope case of the moment result
\cite[Theorem~5 and Lemma~A.2]{Pesenti2025robust} to $g_\lambda$.
The value is $\sigma b_\lambda$, uniquely attained by
the quantile \eqref{(5)} over $\mathcal M(\mu,\sigma)$.
That quantile is symmetric, so the same value and uniqueness hold
on $\mathcal M^s(\mu,\sigma)$. The argument requires
$b_\lambda>0$, as stipulated in the statement.
\end{proof}

\subsection{Proof of Lemma~\ref{le.3}}
\begin{proof}
Write $q_\lambda=(g_\lambda^*)'$ and use $R_F$ from the proof of
Lemma~\ref{lem:sym-envelope-regularity}.
Uniform convergence of $g_\lambda$ as $\lambda\to\lambda_*$ implies
uniform convergence of its concave envelope. Secant-slope bounds
for concave functions then give
$q_\lambda(t)\to q_{\lambda_*}(t)$ at every continuity point of
$q_{\lambda_*}$ in $(0,1)$, hence almost everywhere.
For $0\leq\lambda\leq L$,
\[
0\leq g_\lambda^*\leq\bar g^*+LR_F.
\]
The convex-order comparison in the proof of
Lemma~\ref{lem:sym-envelope-regularity} yields
$q_\lambda(1-U)\leq_{cx}Y_L$, where
$Y_L=(\bar g^*)'(1-U)+LR_F'(1-U)\in L^2$.
For $K>0$, convexity of $x\mapsto(|x|-K)_+^2$ gives
\[
\sup_{0\leq\lambda\leq L}
\mathbb E\!\left[q_\lambda(1-U)^2
 \mathbf1_{\{|q_\lambda(1-U)|>2K\}}\right]
\leq4\mathbb E[(|Y_L|-K)_+^2]\longrightarrow0.
\]
Thus the squared derivatives are uniformly integrable, and the
almost-everywhere convergence implies
$\notationedit{\|q_\lambda-q_{\lambda_*}\|}\to0$.
Consequently, $c$ is continuous wherever $b_{\lambda_*}>0$,
including at zero when $b_0>0$.

For monotonicity, let $J(G)=\rho_{\hat{\bar g}}(G)$ and
$K(G)=\langle F^{-1},G^{-1}-\mu\rangle$.
Replacing $\bar g$ by $\hat{\bar g}$ leaves $g_\lambda^*$ unchanged.
Lemma~\ref{le.4} therefore makes $H_\lambda$ the unique maximizer
of $J(G)+2\lambda K(G)$ over $\mathcal M^s(\mu,\sigma)$.
For $0<\lambda_1<\lambda_2$, adding the two optimality inequalities gives
\[
2(\lambda_2-\lambda_1)
\bigl(K(H_{\lambda_2})-K(H_{\lambda_1})\bigr)\geq0.
\]
Since $K(H_\lambda)=\sigma\sigma_Fc(\lambda)$, $c$ is nondecreasing.
The upper and lower envelope bounds make $J$ bounded on this
moment class. Comparison with $H_\infty$ gives
\[
0\leq2\lambda\bigl(\sigma\sigma_0-K(H_\lambda)\bigr)
\leq J(H_\lambda)-J(H_\infty).
\]
It follows that $K(H_\lambda)\to\sigma\sigma_0$ as
$\lambda\to\infty$. Monotonicity and boundedness also give the
right limit $c_0$; the preceding $L^2$ convergence identifies it
with the correlation at zero only when $b_0>0$.
\end{proof}

\subsection{Proof of Theorem~\ref{th.3}}
\begin{proof}
For part~(i), Lemma~\ref{le.3} and \eqref{eq:symmetric-upper-budget}
provide $\lambda_\epsilon>0$ with
$d_W^2(F,H_{\lambda_\epsilon})=\epsilon$.
Set $K(G)=\langle F^{-1},G^{-1}-\mu\rangle$.
Every feasible $G$ satisfies
$K(G)\geq K(H_{\lambda_\epsilon})$.
Using $\rho_{g_\lambda}(G)=\rho_{\bar g}(G)+2\lambda K(G)$
and Lemma~\ref{le.4}, we obtain
\[
\rho_{\bar g}(G)
\leq\rho_{\bar g}(H_{\lambda_\epsilon})
 +2\lambda_\epsilon\bigl(K(H_{\lambda_\epsilon})-K(G)\bigr)
\leq\rho_{\bar g}(H_{\lambda_\epsilon}).
\]
Equality requires $G=H_{\lambda_\epsilon}$ by uniqueness in the
auxiliary problem. This proves uniqueness of the distribution;
uniqueness of the multiplier is not required.
Proposition~\ref{prop.1} gives the asserted value for $\rho_g$.

For part~(ii), $H_0$ is feasible because
$d_W^2(F,H_0)=\epsilon_1\leq\epsilon$.
Lemma~\ref{le.5} gives its value and uniqueness.

For part~(iii), $\bar g\leq0$ gives the upper bound $\mu g(1)$.
Compare the auxiliary optimizer $H_\lambda$ with $P_n$ in
\eqref{eq:zero-envelope-sequence}. Cauchy--Schwarz gives
$|K(G)|\leq\sigma\sigma_F$ on the moment class, so
\[
0\geq\rho_{\bar g}(H_\lambda)
\geq\rho_{\bar g}(P_n)-4\lambda\sigma\sigma_F.
\]
Let $n\to\infty$ and then $\lambda\downarrow0$.
Every $H_\lambda$ is feasible at budgets $\epsilon\geq\epsilon_1$,
since $c(\lambda)\geq c_0$, and its objective tends to $\mu g(1)$.
The upper bound is therefore sharp even when $H_0$ is undefined.
If $\bar g\equiv0$, Proposition~\ref{prop.1} makes the objective constant.
\end{proof}

\subsection{Proof of Theorem~\ref{th.3.1}}
\begin{proof}
We show that, for every $\epsilon>\epsilon_0$,
\begin{equation}\label{eq:symmetric-usc-relaxation}
\sup_{G\in\mathcal M_\epsilon^s(\mu,\sigma)}\rho_{\bar g}(G)
=\sup_{G\in\mathcal M_\epsilon^s(\mu,\sigma)}\rho_{\hat{\bar g}}(G).
\end{equation}
The inequality from left to right follows from $\bar g\leq\hat{\bar g}$.
For the reverse inequality, fix a feasible $G$ and write
$x=G^{-1}-\mu$, $x_\infty=H_\infty^{-1}-\mu$.
If $G\ne H_\infty$, define, for $0<\theta<1$,
\[
y_\theta=(1-\theta)x+\theta x_\infty,
\qquad
G_\theta^{-1}=\mu+\sigma\frac{y_\theta}{\notationedit{\|y_\theta\|}}.
\]
These are symmetric quantiles with the prescribed moments.
Writing $d=\langle x,x_\infty\rangle/\sigma^2\in[0,1)$, we have
\[
\frac{\langle G_\theta^{-1}-\mu,x_\infty\rangle}{\sigma^2}
=\frac{(1-\theta)d+\theta}
 {\sqrt{(1-\theta)^2+\theta^2+2\theta(1-\theta)d}}>d.
\]
Hence $d_W^2(F,G_\theta)<d_W^2(F,G)\leq\epsilon$.
Comonotonic additivity and positive homogeneity give
\[
\rho_{\hat{\bar g}}(G_\theta)
=\frac{\sigma}{\notationedit{\|y_\theta\|}}
 \bigl((1-\theta)\rho_{\hat{\bar g}}(G)
       +\theta\rho_{\hat{\bar g}}(H_\infty)\bigr)
\longrightarrow\rho_{\hat{\bar g}}(G).
\]
For each fixed $\theta$, apply Lemma~\ref{lem:sym-usc-approximation}
to $G_\theta$. The approximating laws remain feasible for all
sufficiently large indices because the budget inequality is strict.
Their $\rho_{\bar g}$-values approach
$\rho_{\hat{\bar g}}(G_\theta)$.
Letting $\theta\downarrow0$ gives the reverse inequality for $G$.
If $G=H_\infty$, the budget is already strict and the approximation
lemma applies directly. Taking the supremum proves
\eqref{eq:symmetric-usc-relaxation}.

The upper semicontinuous distortion $\hat{\bar g}$ has zero endpoint
values and the same concave envelopes as $\bar g$ after addition
of $\lambda R_F$. Apply Theorem~\ref{th.3} to $\hat{\bar g}$ and
then use Proposition~\ref{prop.1} and
\eqref{eq:symmetric-usc-relaxation}.
This proves both parts, including the case $b_0=0$.
The argument establishes equality of the suprema, not attainment for
the original riskmetric. At $\epsilon=\epsilon_0$, the strict-feasibility
argument no longer applies; Lemma~\ref{lem:sym-feasibility} gives
the value $\rho_g(H_\infty)$ directly.
\end{proof}

\subsection{Proof of Proposition~\ref{th.1}}
\begin{proof}
Write $S=\sigma_0^2$, $V=V_{\bar g}$, $C=C_{\bar g,F}$ and
$D=C_{\epsilon,F}$. Concavity of $g$ implies concavity of $\bar g$
and $g_\lambda$, so no concavification is needed. In $L^2(0,1)$,
\[
(g_\lambda^*)'(1-u)=\bar g'(1-u)+2\lambda s_F(u),\qquad
b_\lambda^2=V-4\lambda C+4\lambda^2S.
\]
Here $\langle F^{-1},\bar g'(1-\cdot)\rangle=-C$ and
$\langle F^{-1},s_F\rangle=S$. Hence the active budget equation is
\[
D=\sigma\frac{-C+2\lambda S}
{\sqrt{V-4\lambda C+4\lambda^2S}}.
\]
When $SV-C^2>0$, its right-hand side is strictly increasing from
$-\sigma C/\sqrt V$ to $\sigma\sqrt S$. The active-budget condition
places $D$ strictly between these endpoints. Setting $x=2\lambda S-C>0$
and squaring yields
\[
x=D\sqrt{\frac{SV-C^2}{\sigma^2S-D^2}},
\]
which gives \eqref{e1}; the endpoint inequalities ensure $\lambda>0$.
If $V>0$ and $SV=C^2$, equality in Cauchy--Schwarz makes
$\bar g'(1-\cdot)$ a positive multiple of $s_F$.
Then $H_\lambda=H_\infty$ for all $\lambda\geq0$ and
$\epsilon_1=\epsilon_0$, so the active interval is empty.
For $V>0$, the inactive case follows from Theorem~\ref{th.3}.
Finally, if $V=0$, then $\bar g=0$. The concave function
$g(t)-tg(1)$ is nonnegative and its values at $t$ and $1-t$ sum to zero;
it therefore vanishes identically. Thus $g(t)=tg(1)$, $H_\lambda=H_\infty$
for $\lambda>0$, and $\epsilon_1=\epsilon_0$. Every feasible distribution
then has risk $\mu g(1)$, as stated.
\end{proof}

\subsection{Proof of Corollary~\ref{ex.5}}
\begin{proof}
Fix $\lambda=\lambda_\epsilon>0$, and write
$c=1-\beta$, $a=1-\alpha$ and $d=1/[2(\beta-\alpha)]$.
The function $r_\lambda$ is concave and nondecreasing on $[0,1/2]$.
The functions $A_\lambda$ and $B_\lambda$ are continuous, and
$A_\lambda(m)-B_\lambda(m)$ is nondecreasing: increasing $m$ by $h>0$
decreases $A_\lambda$ by at most $ch$ and $B_\lambda$ by at least $ch$.
Since $0<c<a$, their difference tends to $-\infty$ and $+\infty$
as $m$ tends to $-\infty$ and $+\infty$, respectively.
Its zero set is therefore nonempty and compact, so $m_\lambda$ is well defined.

The common supporting line $\ell_\lambda$ majorizes both branches
and touches them at $u_\lambda\leq c\leq v_\lambda\leq a$.
Replacing $g_\lambda$ by this line on $[u_\lambda,v_\lambda]$
preserves concavity at the contacts. On $[a,1/2]$, the function is
$1/2+r_\lambda$, with a downward slope change at $a$.
Thus the replacement is concave and nondecreasing on $[0,1/2]$;
reflection about $1/2$ gives a concave majorant on $[0,1]$.
Every concave majorant must dominate the chord between the two contacts,
and hence this construction is the least concave majorant.
The leftmost and rightmost contact convention includes any affine contact
intervals. If the contacts coincide, no replacement is needed; any
nonuniqueness of supporting slopes or contacts leaves the envelope unchanged.

Its right derivative, reflected as in \eqref{(5)}, gives the displayed
left-continuous quantile. On $[c,a]$ the envelope has slope $m_\lambda$
up to $v_\lambda$ and slope $d+r_\lambda'$ thereafter. Consequently,
\[
\int_c^a(g_\lambda^*)'(s)\,ds
=m_\lambda(v_\lambda-c)+d(a-v_\lambda)
+\lambda\int_{v_\lambda}^{a}
\bigl[F^{-1}(1-s)-F^{-1}(s)\bigr]\,ds.
\]
Substitution into
$\RVaR_{\alpha,\beta}(H_\lambda)
=\mu+\sigma[(\beta-\alpha)b_\lambda]^{-1}
\int_c^a(g_\lambda^*)'(s)\,ds$
proves the formula. Optimality and uniqueness follow from
Theorem~\ref{th.3}.
\end{proof}

\subsection{Proof of Corollary~\ref{ex.6}}
\begin{proof}
The function $R_F$ is continuous, concave and symmetric about $1/2$.
Since taking the upper semicontinuous modification leaves the
concave envelope unchanged, we work with
$\hat{\bar g}+\lambda R_F$.

For $\alpha>1/2$, set $a=1-\alpha$,
$\tau=t_{\alpha,\lambda}$, $m_+=m_{\alpha,\lambda}^+$, and
$A_+=1/2+\lambda R_F(a)$. Here
$\hat{\bar g}(t)=\tfrac12\mathbf1_{[a,1-a]}(t)$, and
\begin{equation}\label{eq:var-upper-envelope}
g_\lambda^*(t)=
\begin{cases}
\lambda R_F(t),&t\in[0,\tau)\cup(1-\tau,1],\\
A_++m_+(t-a),&t\in[\tau,a),\\
1/2+\lambda R_F(t),&t\in[a,1-a],\\
A_++m_+(1-a-t),&t\in(1-a,1-\tau].
\end{cases}
\end{equation}
The minimizing property in \eqref{eq:var-upper-contact} ensures
that the left affine segment majorizes $\lambda R_F$ and joins
it at $\tau$. It also gives the one-sided slope inequalities
$\lambda\partial_+R_F(\tau)\leq m_+
\leq\lambda\partial_-R_F(\tau)$ when $\tau>0$; at $\tau=0$
only the first inequality is needed. Thus the joins are concave.
Symmetry gives the right segment. Every concave majorant lies
above the chords between these contact points, so
\eqref{eq:var-upper-envelope} is the least such majorant.

For $\alpha<1/2$, we have
$\hat{\bar g}(t)=-\tfrac12\mathbf1_{(\alpha,1-\alpha)}(t)$.
Set $A_0=\lambda R_F(\alpha)$.
If $\Delta_{\alpha,\lambda}\leq1/2$, the central part of
$\hat{\bar g}+\lambda R_F$ never exceeds $A_0$, and
\begin{equation}\label{eq:var-flat-envelope}
g_\lambda^*(t)=
\begin{cases}
\lambda R_F(t),&t\in[0,\alpha)\cup(1-\alpha,1],\\
A_0,&t\in[\alpha,1-\alpha].
\end{cases}
\end{equation}
This function is concave. Every concave majorant lies above its
constant segment because the contact values at $\alpha$ and
$1-\alpha$ are equal. In this case, \eqref{eq:var-flat-envelope} shows that
$g_\lambda^*/\lambda$ is independent of $\lambda$.
The same expression holds for every smaller positive multiplier,
so the normalized quantile $h_\lambda$ is constant along this part
of the multiplier path. Consequently, $c(\lambda)=c_0$ and
$d_W^2(F,H_\lambda)=\epsilon_1$.
Thus the flat-envelope case cannot occur when
$d_W^2(F,H_\lambda)=\epsilon<\epsilon_1$; throughout the active range
with $\alpha<1/2$, we must have $\Delta_{\alpha,\lambda}>1/2$.

If $\Delta_{\alpha,\lambda}>1/2$, put
$\tau=\tilde t_{\alpha,\lambda}$ and $m_-=m_{\alpha,\lambda}^-$.
Then
\begin{equation}\label{eq:var-lower-envelope}
g_\lambda^*(t)=
\begin{cases}
\lambda R_F(t),&t\in[0,\alpha)\cup(1-\alpha,1],\\
A_0+m_-(t-\alpha),&t\in[\alpha,\tau),\\
-1/2+\lambda R_F(t),&t\in[\tau,1-\tau],\\
A_0+m_-(1-\alpha-t),&t\in(1-\tau,1-\alpha].
\end{cases}
\end{equation}
The affine segments join the central curve because
\eqref{eq:var-lower-contact} gives
\[
A_0+m_-(\tau-\alpha)=-1/2+\lambda R_F(\tau),
\]
so the pieces agree at $\tau$ and, by symmetry, at $1-\tau$.
The maximizing property makes the affine segments majorize the
central curve and ensures the one-sided slope inequalities at
the joins. Together with concavity of $R_F$, this proves concavity
of \eqref{eq:var-lower-envelope}. On the remaining intervals it
coincides with $\hat{\bar g}+\lambda R_F$; the chord argument again
proves minimality. These arguments also cover nondifferentiable
contact points and $\tau=1/2$.

Finally, the upper semicontinuous symmetrized risk equals
\[
\rho_{\hat{\bar g}}(H_\lambda)=
\begin{cases}
\tfrac12\bigl(h_\lambda(\alpha+)-h_\lambda(1-\alpha)\bigr),
 &\alpha>1/2,\\
-\tfrac12\bigl(h_\lambda(1-\alpha)-h_\lambda(\alpha+)\bigr),
 &\alpha<1/2.
\end{cases}
\]
In the active range, the corresponding envelope slopes give
$\sigma m_+/b_\lambda$ and $-\sigma m_-/b_\lambda$, respectively.
Theorem~\ref{th.3.1} now gives \eqref{eq:var-value}.

It remains to prove nonattainment for the left quantile.
The symmetrization of $g_\alpha^+(t)=\mathbf1_{[1-\alpha,1]}(t)$
is $\hat{\bar g}$. Thus Theorems~\ref{th.3} and~\ref{th.3.1}
identify $H_\lambda$ as the unique maximizer of $\VaR_\alpha^+$,
whose maximum equals the supremum of $\VaR_\alpha$.
Since $\VaR_\alpha(G)\leq\VaR_\alpha^+(G)$ for every $G$,
any maximizer of $\VaR_\alpha$ would have to be $H_\lambda$.

If $\alpha>1/2$, let $a=1-\alpha$ and $\tau=t_{\alpha,\lambda}<a$.
By \eqref{eq:var-upper-contact} and concavity of $R_F$,
\[
m_+
=\frac{1}{2(a-\tau)}
 +\lambda\frac{R_F(a)-R_F(\tau)}{a-\tau}
>\lambda\partial_+R_F(a).
\]
The left-continuous quantile obtained from
\eqref{eq:var-upper-envelope} therefore satisfies
\[
\VaR_\alpha(H_\lambda)
=\mu+\frac{\sigma\lambda}{b_\lambda}\partial_+R_F(a)
<\mu+\frac{\sigma m_+}{b_\lambda}.
\]
If $\alpha<1/2$, the active range has
$\Delta_{\alpha,\lambda}>1/2$. Set
$\tau=\tilde t_{\alpha,\lambda}>\alpha$.
Equation~\eqref{eq:var-lower-contact} and concavity of $R_F$ give
\[
m_-
=\lambda\frac{R_F(\tau)-R_F(\alpha)}{\tau-\alpha}
 -\frac{1}{2(\tau-\alpha)}
<\lambda\partial_-R_F(\alpha).
\]
Using the left-continuous quantile from
\eqref{eq:var-lower-envelope}, we obtain
\[
\VaR_\alpha(H_\lambda)
=\mu-\frac{\sigma\lambda}{b_\lambda}\partial_-R_F(\alpha)
<\mu-\frac{\sigma m_-}{b_\lambda}.
\]
These strict inequalities exclude the only possible maximizing
law and prove nonattainment throughout the active range.

Finally, suppose $\alpha>1/2$, $\epsilon\geq\epsilon_1$ and
$\epsilon>\epsilon_0$.
Theorem~\ref{th.3} applied to $g_\alpha^+$ gives the unique maximizer
$H_0$ of $\VaR_\alpha^+$, with value
$\mu+\sigma/\sqrt{2(1-\alpha)}$.
The same value is the left-VaR supremum by
Theorem~\ref{th.3.1}, whereas the left-continuous quantile of $H_0$
satisfies $\VaR_\alpha(H_0)=\mu$.
Hence this supremum is also unattained, which proves the asserted
extension to all $\epsilon>\epsilon_0$ for $\alpha>1/2$.
\end{proof}

\section{Proofs of the main results in Section~\ref{sec4}}\label{app:unimodality}
\subsection{Proof of Lemma~\ref{lem:us-cone}}
\begin{proof}
A centered symmetric unimodal quantile is the antisymmetric extension
of a nonnegative, nondecreasing convex function on $[1/2,1)$ that
vanishes at $1/2$. These properties are preserved by addition and
nonnegative scaling, so $\mathcal C$ is a convex cone.
To check closedness, let $z_n\in\mathcal C$ converge to $z$ in $L^2$.
Antisymmetry and zero mean pass to the limit. Nonnegativity,
monotonicity and convexity on $(1/2,1)$ also pass to the limit;
for the last two properties, the distributional derivatives retain
their signs. Choose the resulting convex representative on that interval.
For $1/2<u<v<1$, convexity gives
\[
0\leq z_n(u)\leq\frac{u-1/2}{v-1/2}z_n(v).
\]
Passing along an almost everywhere convergent subsequence gives the same
inequality for $z$, first almost everywhere and then everywhere by
continuity. In particular, $z(u)\to0$ as $u\downarrow1/2$.
Antisymmetric extension is therefore continuous at $1/2$, and
Lemma~\ref{le.a1} implies $z\in\mathcal C$.
Since $\mathcal M^{us,-1}=\mathbb R\mathbf1+\mathcal C$ is an orthogonal
sum, it too is a closed convex cone.

Let $f^{\mathrm s}(u)=[f(u)+f(1-u)]/2$.
Every $q\in\mathcal M^{us,-1}$ has a unique representation $q=b+z$
with $b\in\mathbb R$ and $z\in\mathcal C$.
Orthogonality of symmetric and antisymmetric functions gives
\begin{align}\label{eq:decomposition}
\notationedit{\|f-(b+z)\|}^2
&=\notationedit{\|f^{\mathrm a}-z\|}^2+\notationedit{\|f^{\mathrm s}-b\|}^2\nonumber\\
&=\notationedit{\|f^{\mathrm a}-z\|}^2+\notationedit{\|f^{\mathrm s}-A_f\|}^2+(b-A_f)^2.
\end{align}
Minimizing separately in $b$ and $z$ proves \eqref{eq:generalidentity}.
\end{proof}

\subsection{Proof of Lemma~\ref{le.6a}}
\begin{proof}
For $G^{-1}\in\mathcal{M}^{us,-1}(\mu,\sigma)$,
\begin{align*}
 d_{W}^{2}(G^{-1},F^{-1})=(\mu_{F}-\mu)^2+(\sigma_{F}-\sigma)^2+2\mu\mu_{F}+2\sigma\sigma_{F}-2\langle G^{-1},F^{-1}\rangle.
 \end{align*}
By Theorem~2.2 of \cite{brunk1965conditional}, for
\[
G^{-1}\in\mathcal{M}^{us,-1}(\mathbb{E}(F^{-1,\uparrow}(U)),\sqrt{\mathrm{Var}(F^{-1,\uparrow}(U))})
,
\]

\begin{align*}
 \langle F^{-1}-F^{-1,\uparrow},F^{-1,\uparrow}-G^{-1}\rangle \geq0.
 \end{align*}
and therefore
\begin{align*}
 \langle F^{-1},G^{-1}\rangle \leq \langle F^{-1},F^{-1,\uparrow}\rangle +\langle F^{-1,\uparrow},G^{-1}\rangle-\langle F^{-1,\uparrow},F^{-1,\uparrow}\rangle.
 \end{align*}
The Cauchy--Schwarz inequality and the definition of the $L^{2}$-projection give

\[
\langle F^{-1,\uparrow},G^{-1}\rangle\leq\langle F^{-1,\uparrow},F^{-1,\uparrow}\rangle.
\]

Hence, for $G^{-1}\in\mathcal{M}^{us,-1}(\mathbb{E}(F^{-1,\uparrow}(U)),\sqrt{\mathrm{Var}(F^{-1,\uparrow}(U))})$,
$\langle F^{-1},G^{-1}\rangle\leq\langle F^{-1},F^{-1,\uparrow}\rangle$. Rescaling to the prescribed moments shows that, for $G^{-1}\in\mathcal{M}^{us,-1}(\mu,\sigma)$,
\[
\langle F^{-1},G^{-1}\rangle\leq\langle F^{-1},\mu+\sigma\frac{F^{-1,\uparrow}-\mathbb{E}(F^{-1,\uparrow}(U))}{\sqrt{\mathrm{Var}(F^{-1,\uparrow}(U))}}\rangle
.
\]
 Consequently, for $G^{-1}\in\mathcal{M}^{us,-1}(\mu,\sigma)$,
\begin{align}\label{(c1)}
 \nonumber d_{W}^{2}(G^{-1},F^{-1})&\geq(\mu_{F}-\mu)^2+(\sigma_{F}-\sigma)^2+2\mu\mu_{F}+2\sigma\sigma_{F}-2\langle F^{-1},\mu+\sigma\frac{F^{-1,\uparrow}-\mathbb{E}(F^{-1,\uparrow}(U))}{\sqrt{\mathrm{Var}(F^{-1,\uparrow}(U))}}\rangle\\
 \nonumber&=(\mu_{F}-\mu)^2+(\sigma_{F}-\sigma)^2+2\sigma\sigma_{F}(1-\mathrm{Corr}(F^{-1}(U),F^{-1,\uparrow}(U)))\\
 &=(\mu_{F}-\mu)^2+(\sigma_{F}-\sigma)^2+2\sigma\sigma_{F}(1-\bar{c}_{0}).
 \end{align}
Equality in the Cauchy--Schwarz step forces the rescaled quantile
to coincide almost everywhere with $F^{-1,\uparrow}$, because the
two functions have the same mean and $L^2$ norm.
Thus equality in \eqref{(c1)} is possible only for the displayed
standardized projection, which itself attains equality. It follows from \eqref{(c1)} that, if $\epsilon<(\mu_{F}-\mu)^2+(\sigma_{F}-\sigma)^2+2\sigma\sigma_{F}(1-\bar{c}_{0})$, then $\mathcal{M}^{us,-1}_{\epsilon}(\mu,\sigma)=\emptyset$. If $\epsilon=(\mu_{F}-\mu)^2+(\sigma_{F}-\sigma)^2+2\sigma\sigma_{F}(1-\bar{c}_{0})$, then $\mathcal{M}^{us,-1}_{\overline{\epsilon}_0}(\mu,\sigma)=\left\{\mu+\sigma\frac{F^{-1,\uparrow}-\mathbb{E}(F^{-1,\uparrow}(U))}{\sqrt{\mathrm{Var}(F^{-1,\uparrow}(U))}}\right\}$.
\end{proof}

\subsection{Proof of Proposition~\ref{le.6}}
\begin{proof}
Fix $\lambda\geq0$ with $b_{\lambda,us}>0$. Then $h_{\lambda,us}^{\uparrow}\in\mathcal{M}^{us,-1}(\mu,\sigma)$. Given any $h\in\mathcal{M}^{us,-1}(\mu,\sigma)$, the function $k(u)=b_{\lambda,us}\frac{h(u)-\mu}{\sigma}+a_{\lambda,us}$ is a symmetric unimodal quantile. Moreover, $\|
 k_{\lambda,us}^{\uparrow}\|^{2}=\|
 k\|^{2}$. By the definition of $k_{\lambda,us}^{\uparrow}$,
\begin{align*}
 \| k_{\lambda}(u)-k_{\lambda,us}^{\uparrow}(u)\|^{2}\leq\| k_{\lambda}(u)-k(u)\|^{2}\\
 \Leftrightarrow\langle k_{\lambda}(u),k_{\lambda,us}^{\uparrow}(u)\rangle\geq\langle k_{\lambda}(u),k(u)\rangle\\
 \Leftrightarrow\langle k_{\lambda}(u),h_{\lambda,us}^{\uparrow}(u)\rangle\geq\langle k_{\lambda}(u),h(u)\rangle.
 \end{align*}
Equality holds if and only if $h(u)=h_{\lambda,us}^{\uparrow}(u)$, establishing uniqueness.

For the remaining assertions, assume the nondegeneracy conditions in the second part of the proposition.
Projection nonexpansiveness
\citep[Theorem~2.3]{brunk1965conditional} gives
$||k_{\lambda_{2},us}^{\uparrow}(u)-k_{\lambda_{1},us}^{\uparrow}(u)||_{2}^{2}\leq(\lambda_{2}-\lambda_{1})^{2}(\mu_{F}^{2}+\sigma_{F}^{2})$.
Since the projected variances are positive, the correlation
$\mathrm{Corr}(F^{-1}(U),h_{\lambda,us}^{\uparrow}(U))
=\mathrm{Corr}(F^{-1}(U),k_{\lambda,us}^{\uparrow}(U))$
is continuous. To find its limit as $\lambda\to\infty$, define
\[
l_{\lambda,us}(u)
:=\frac{k_{\lambda}(u)}{\lambda}
=F^{-1}(u)+\frac{\bar{g}'(1-u)}{\lambda},
\qquad
l_{\infty}=F^{-1},
\qquad
l_{\infty}^{\uparrow}=F^{-1,\uparrow}.
\]
Since $\mathcal{M}^{us,-1}$ is a cone, its projection is positively
homogeneous. Thus
$l_{\lambda,us}^{\uparrow}=k_{\lambda,us}^{\uparrow}/\lambda$
is the projection of $l_{\lambda,us}$. By nonexpansiveness in
$L^{2}(0,1)$ (Theorem~2.3 of \cite{brunk1965conditional}),
\[
\bigl\|l_{\lambda,us}^{\uparrow}-l_{\infty}^{\uparrow}\bigr\|
\leq
\bigl\|l_{\lambda,us}-l_{\infty}\bigr\|
=\frac{\|\bar{g}'\|}{\lambda}
\longrightarrow0
\qquad\text{as }\lambda\longrightarrow\infty.
\]
This $L^{2}$ convergence and the Cauchy--Schwarz inequality imply
\begin{align*}
\mathbb{E}\bigl[l_{\lambda,us}^{\uparrow}(U)\bigr]
&\longrightarrow
\mathbb{E}\bigl[l_{\infty}^{\uparrow}(U)\bigr],\\
\mathbb{E}\bigl[(l_{\lambda,us}^{\uparrow}(U))^{2}\bigr]
&\longrightarrow
\mathbb{E}\bigl[(l_{\infty}^{\uparrow}(U))^{2}\bigr],\\
\mathbb{E}\bigl[F^{-1}(U)l_{\lambda,us}^{\uparrow}(U)\bigr]
&\longrightarrow
\mathbb{E}\bigl[F^{-1}(U)l_{\infty}^{\uparrow}(U)\bigr].
\end{align*}
In particular, the variances converge to
$\mathrm{Var}(F^{-1,\uparrow}(U))>0$, since $F^{-1,\uparrow}$
is nonconstant by assumption. Positive scaling leaves correlation
unchanged, so
\begin{equation*}
\lim_{\lambda\rightarrow\infty}
\mathrm{Corr}\bigl(F^{-1}(U),k_{\lambda,us}^{\uparrow}(U)\bigr)
=
\lim_{\lambda\rightarrow\infty}
\mathrm{Corr}\bigl(F^{-1}(U),l_{\lambda,us}^{\uparrow}(U)\bigr)=\mathrm{Corr}\bigl(F^{-1}(U),F^{-1,\uparrow}(U)\bigr).
\end{equation*}

To prove monotonicity, set
$J(G)=\rho_{\bar g}(G)$ and
$K(G)=\langle F^{-1},G^{-1}-\mu\rangle$.
The auxiliary objective is $J(G)+\lambda K(G)$, up to a constant
common to the moment class. For $0<\lambda_1<\lambda_2$, add its
two optimality inequalities to obtain
\[
(\lambda_2-\lambda_1)
\bigl(K(H_{\lambda_2,us}^{\uparrow})
      -K(H_{\lambda_1,us}^{\uparrow})\bigr)\geq0.
\]
Since $K(H_{\lambda,us}^{\uparrow})=\sigma\sigma_Fc_{us}(\lambda)$,
the correlation is nondecreasing and bounded, and $c_1$ exists.
Nonexpansiveness of the projection also gives
$\notationedit{\|k_{\lambda,us}^{\uparrow}-k_{0,us}^{\uparrow}\|}
\leq\lambda\notationedit{\|F^{-1}\|}$.
If $b_{0,us}>0$, normalization is continuous at zero, which
identifies $c_1$ with the stated correlation.
Finally,
\[
d_W^2(F,H_{\lambda,us}^{\uparrow})
=(\mu_F-\mu)^2+\sigma_F^2+\sigma^2
 -2\sigma\sigma_Fc_{us}(\lambda)
\]
yields the distance monotonicity and both budget limits.
\end{proof}

\subsection{Proof of Theorem~\ref{th.4}}
\begin{proof}
Write $J(G)=\rho_{\bar g}(G)$ and
$K(G)=\langle F^{-1},G^{-1}-\mu\rangle$.
For part~(i), continuity and the two limits in
Proposition~\ref{le.6} provide a $\lambda_\epsilon>0$ such that
$d_W^2(F,H_{\lambda_\epsilon,us}^{\uparrow})=\epsilon$.
Every feasible $G$ satisfies
$K(G)\geq K(H_{\lambda_\epsilon,us}^{\uparrow})$.
Optimality for $J+\lambda_\epsilon K$ gives
\begin{equation*}
J(G)
\leq J(H_{\lambda_\epsilon,us}^{\uparrow})
 +\lambda_\epsilon\bigl(K(H_{\lambda_\epsilon,us}^{\uparrow})-K(G)\bigr)\leq J(H_{\lambda_\epsilon,us}^{\uparrow}).
\end{equation*}
Equality requires $G=H_{\lambda_\epsilon,us}^{\uparrow}$ by
uniqueness in Proposition~\ref{le.6}. Proposition~\ref{prop.1}
then yields the asserted value and uniqueness for $\rho_g$.

For part~(ii), first suppose $b_{0,us}>0$.
Proposition~\ref{le.6} implies
$d_W^2(F,H_{0,us}^{\uparrow})=\overline{\epsilon}_1\leq\epsilon$.
The same proposition at $\lambda=0$ gives uniqueness over the
full moment-and-shape class, hence also over the Wasserstein-constrained class.

If $b_{0,us}=0$, the projection $k_{0,us}^{\uparrow}$ is zero:
projection preserves the mean, and
$\int_0^1\bar g'(1-u)\,\mathrm du=0$.
The projection inequality gives $J(G)\leq0$ on
$\mathcal M^{us}(\mu,\sigma)$.
For $n\geq1$, consider
\[
G_n=(1-1/n)\delta_\mu
 +\frac1n\operatorname{Unif}
 [\mu-\sqrt{3n}\sigma,\mu+\sqrt{3n}\sigma].
\]
These laws are symmetric and unimodal with the prescribed moments,
and their quantiles equal $\mu$ on $[1/(2n),1-1/(2n)]$.
Cauchy--Schwarz therefore yields
\[
|J(G_n)|\leq\sigma\left(
\int_0^{1/(2n)}[\bar g'(1-u)]^2\,\mathrm du
+\int_{1-1/(2n)}^1[\bar g'(1-u)]^2\,\mathrm du
\right)^{1/2}\longrightarrow0.
\]
Auxiliary optimality and $|K(G)|\leq\sigma\sigma_F$ give
\[
0\geq J(H_{\lambda,us}^{\uparrow})
\geq J(G_n)-2\lambda\sigma\sigma_F.
\]
Letting $n\to\infty$ and then $\lambda\downarrow0$ shows that
$J(H_{\lambda,us}^{\uparrow})\to0$.
These laws are feasible for every $\lambda>0$ when
$\epsilon\geq\overline{\epsilon}_1$, because the distance is
nonincreasing and has right limit $\overline{\epsilon}_1$ at zero.
Proposition~\ref{prop.1} therefore gives the supremum $\mu g(1)$,
whether or not it is attained.

It remains to verify the explicit value and the dual representation.
Set
\[
p_\lambda=P_{\mathcal C}k_\lambda^{\mathrm a}
=k_{\lambda,us}^{\uparrow}-a_{\lambda,us}.
\]
Projection onto the closed convex cone $\mathcal C$ gives
$\langle k_\lambda^{\mathrm a}-p_\lambda,p_\lambda\rangle=0$
and
$\langle k_\lambda^{\mathrm a}-p_\lambda,z\rangle\leq0$
for every $z\in\mathcal C$.
Consequently, for $G\in\mathcal M^{us}(\mu,\sigma)$,
\[
J(G)+\lambda K(G)
=\langle k_\lambda^{\mathrm a},G^{-1}-\mu\rangle
\leq\sigma\notationedit{\|p_\lambda\|}
=\sigma b_{\lambda,us}.
\]
If $b_{\lambda,us}>0$, equality holds at
$G=H_{\lambda,us}^{\uparrow}$.
For finite $\epsilon$, every feasible $G$ satisfies
$K(G)\geq D_\epsilon$. Thus, for every $\lambda\geq0$,
\[
V_{us}(\epsilon)
\leq\mu g(1)+\sigma b_{\lambda,us}-\lambda D_\epsilon.
\]
In part~(i), $K(H_{\lambda_\epsilon,us}^{\uparrow})=D_\epsilon$
and the auxiliary bound is attained. This proves
\eqref{eq:us-sharp-value} and equality in \eqref{eq:us-dual}
at $\lambda_\epsilon$.
In part~(ii), the value already established equals
$\mu g(1)+\sigma b_{0,us}$, including when $b_{0,us}=0$.
The preceding bound at $\lambda=0$ is therefore exact, proving
the remaining dual assertion.
\end{proof}

\subsection{Proof of Corollary~\ref{ex.7}}
\begin{proof}
Here $\bar g_r(t)=rt(1-t)$ and
$\bar g_r'(1-u)=r(2u-1)$.
Because $F\in\mathcal M^{us}$ and $r,\lambda\geq0$, the function
$k_\lambda(u)=r(2u-1)+\lambda F^{-1}(u)$ belongs to the closed
convex cone $\mathcal M^{us,-1}$. Its projection is itself, its
mean is $\lambda\mu_F$, and its variance is
$V_r+2\lambda C_{r,F}+\lambda^2\sigma_F^2$.
Theorem~\ref{th.4} therefore gives \eqref{eq:gdp-quantile}, and
integration against $r(2u-1)$ gives \eqref{eq:gdp-value}.

The projection identities give $\bar c_0=1$ and
$c_1=C_{r,F}/(\sigma_F\sqrt{V_r})$, yielding the stated thresholds.
The candidate quantile satisfies
\[
d_W^2(F,H_{\lambda,us}^{\uparrow})
=(\mu-\mu_F)^2+\sigma^2+\sigma_F^2
 -2\sigma\frac{C_{r,F}+\lambda\sigma_F^2}
 {\sqrt{V_r+2\lambda C_{r,F}+\lambda^2\sigma_F^2}}.
\]
In the active-budget interval,
$\sigma C_{r,F}/\sqrt{V_r}<D_\epsilon<\sigma\sigma_F$ and
$C_{r,F}^2<\sigma_F^2V_r$. Solving
\[
\sigma\frac{C_{r,F}+\lambda\sigma_F^2}
 {\sqrt{V_r+2\lambda C_{r,F}+\lambda^2\sigma_F^2}}
=D_\epsilon
\]
for its positive root yields \eqref{eq:gdp-multiplier}.
The left-hand side is strictly increasing in $\lambda$ in this
case, so the root is unique.
If $C_{r,F}^2=\sigma_F^2V_r$, equality in Cauchy--Schwarz makes
$F^{-1}-\mu_F$ proportional to $2u-1$, and the two thresholds
coincide. Finally, the zero-multiplier quantile is uniform on
$[\mu-\sqrt3\sigma,\mu+\sqrt3\sigma]$ and has value
$\mu+\sigma r/\sqrt3$. Proposition~\ref{le.6} at $\lambda=0$
makes it the unique maximizer over the full moment-and-shape class;
its feasibility for $\epsilon\geq\overline{\epsilon}_1$ proves the
last assertion, including the coincident-threshold case.
\end{proof}

\subsection{Proof of Proposition~\ref{pro.4}}
\begin{proof}
By Lemma~\ref{lem:us-cone},
$P_{us}L_n=a_{L_n}+P_{\mathcal C}L_n^{\mathrm a}$, so the projection
has mean $a_{L_n}$.

To establish the piecewise linear structure, let $\mathcal F_{U,1/2}^{-1}$ denote
the closed convex cone of unimodal quantile functions that are
concave on $(0,1/2)$ and convex on $(1/2,1)$, without a symmetry
restriction. Define
$
p_n=P_{\mathcal F_{U,1/2}^{-1}}L_n^{\mathrm a}.
$
The linear isometry $(Tq)(u)=-q(1-u)$ maps $\mathcal F_{U,1/2}^{-1}$ onto itself and satisfies
$TL_n^{\mathrm a}=L_n^{\mathrm a}$. Hence
$$
\|{L_n^{\mathrm a}-Tp_n}\|=\|{L_n^{\mathrm a}-p_n}\|.
$$
Uniqueness of the projection gives $Tp_n=p_n$. Thus $p_n$ is
antisymmetric and centered, and therefore belongs to $\mathcal{M}^{us,-1}$.
Since $\mathcal{M}^{us,-1}\subset\mathcal F_{U,1/2}^{-1}$, we obtain $p_n=P_{us}L_n^{\mathrm a}$. Equation~\eqref{eq:generalidentity}
then yields
\begin{equation}\label{eq:reconstruction}
L_{n,us}^{\uparrow}=a_{L_n}+p_n.
\end{equation}

The symmetric partition gives
$$
L_n^{\mathrm a}(u)=\eta_{n,i},
\qquad u\in(x_{n,i-1},x_{n,i}),\quad i>n.
$$
By Proposition~3 of \cite{liu2025robust}, applied with transition
point $1/2$, the projection $p_n$ has at most two linear pieces
on each partition interval. Its slopes on the upper half-interval
are nonnegative and nondecreasing, and hence admit the
parameterization in $\mathcal D_n$. Antisymmetry gives $p_n'(u)=p_n'(1-u)$ almost
everywhere, proving \eqref{eq:derivative}. Continuity and
antisymmetry of $p_n$ imply $p_n(1/2)=0$. Integration over $[1/2,1]$,
followed by \eqref{eq:reconstruction}, proves \eqref{eq:endpoint}.
The piecewise linear representation also ensures that the endpoint
value is finite.

Conversely, any parameters in $\mathcal D_n$, together with
\eqref{eq:derivative}--\eqref{eq:endpoint}, determine a continuous
nondecreasing function that is concave on $(0,1/2)$, convex on $(1/2,1)$,
and symmetric about $a_{L_n}$.
The parameterization therefore contains the projection and
only admissible functions.

For a given choice of parameters, denote the resulting function by
$q_n$ and write $z_n=q_n-a_{L_n}$. By antisymmetry of $L_n^{\mathrm a}$ and $z_n$,
\begin{equation}\label{eq:halferror}
\|{L_n^{\mathrm a}-z_n}\|^2
=2\sum_{i=n+1}^{2n}\int_{x_{n,i-1}}^{x_{n,i}}
\bigl(z_n(u)-\eta_{n,i}\bigr)^2\,du.
\end{equation}
The residuals at $x_{n,i-1}$, $\theta_{n,i}$, and $x_{n,i}$ are $e_{n,i}^0$,
$e_{n,i}^1$, and $e_{n,i}^2$, respectively. Integrating the square of the
affine residual on each subinterval gives
\begin{align*}
&\int_{x_{n,i-1}}^{x_{n,i}}\bigl(z_n(u)-\eta_{n,i}\bigr)^2\,du
=\frac{\ell_{n,i}^-}{3}
\big[(e_{n,i}^0)^2+e_{n,i}^0e_{n,i}^1+(e_{n,i}^1)^2\big]
+\frac{\ell_{n,i}^+}{3}
\big[(e_{n,i}^1)^2+e_{n,i}^1e_{n,i}^2+(e_{n,i}^2)^2\big].
\end{align*}
Substituting into \eqref{eq:halferror}, using \eqref{eq:decomposition},
and omitting the common positive factor $2/3$ yields
\eqref{eq:optimization}. Uniqueness follows from the projection theorem.
\end{proof}

\subsection{Proof of Proposition~\ref{prop:fixed-node}}
\begin{proof}
Nonnegative hinge coefficients give nonnegative, nondecreasing slopes
on $[1/2,1]$ and, by antisymmetric extension, an element of
$\mathcal C$. Conversely, every such piecewise linear function has
this representation. Expansion of
$\frac12\notationedit{\|\sum_jc_j\psi_j-L^{\mathrm a}\|}^2$ gives
\eqref{eq:fixed-node-qp}. Linear independence of the hinges makes
$Q$ positive definite, and the stated KKT conditions characterize its
unique minimizer.

The union of the nested cones $\mathcal C_m$ is dense in $\mathcal C$.
To see this, restrict a member of $\mathcal C$ to $[1/2,1-\delta]$
and extend it linearly to $1$ using its left derivative at
$1-\delta$. On the upper half this gives a bounded Lipschitz convex
function between zero and the original function. As $\delta\downarrow0$,
the antisymmetric extensions converge in $L^2$ by dominated convergence.
Linear interpolation of each such function on the fixed partitions
preserves its slope inequalities and converges uniformly. This proves
density. If $p=P_{\mathcal C}L^{\mathrm a}$ and
$v_m\in\mathcal C_m$ tends to $p$, projection optimality gives
\[
\notationedit{\|L^{\mathrm a}-p\|}^2+\notationedit{\|p_m-p\|}^2
\leq\notationedit{\|L^{\mathrm a}-p_m\|}^2
\leq\notationedit{\|L^{\mathrm a}-v_m\|}^2.
\]
Thus $p_m\to p$. Continuity of normalization at $p\ne0$ and the
Cauchy--Schwarz inequality give the remaining assertions.
\end{proof}

\subsection{Proof of Proposition~\ref{pro.5}}
\begin{proof}
By Theorem~2.3 of \cite{brunk1965conditional}, the metric projection
onto a nonempty closed convex subset of $L^2(0,1)$ is nonexpansive.
Applying this result to $\mathcal{M}^{us,-1}$ gives \eqref{eq:projectionerror}.
Define
$$
z=L_{us}^{\uparrow}-a_L,\qquad
z_n=L_{n,us}^{\uparrow}-a_{L_n}.
$$
By \eqref{eq:generalidentity}, $z=P_{us}L^{\mathrm a}$ and $z_n=P_{us}L_n^{\mathrm a}$. Nonexpansiveness
applied to $L_n^{\mathrm a}$ and $L^{\mathrm a}$ yields
$$
\|{z_n-z}\|
\le \|{L_n^{\mathrm a}-L^{\mathrm a}}\|
=\delta_n,
$$
which proves \eqref{eq:centerederror}. Since $b_L=\|{z}\|$ and $b_{L_n}=\|{z_n}\|$,
the reverse triangle inequality gives
\begin{equation}\label{eq:berror}
|b_{L_n}-b_L|
\le \|{z_n-z}\|
\le \delta_n.
\end{equation}
Consequently,
\begin{equation*}
\left\|\frac{z_n}{b_{L_n}}-\frac{z}{b_L}\right\|
\le \frac{\|{z_n-z}\|}{b_L}
+\|{z_n}\|
\left|\frac{1}{b_{L_n}}-\frac{1}{b_L}\right|=\frac{\|{z_n-z}\|+|b_{L_n}-b_L|}{b_L}\le \frac{2\delta_n}{b_L}.
\end{equation*}
Multiplying by $\sigma$ proves \eqref{eq:quantileerror}.

Since $h_{L_n}$ and $h_L$ have identical first two moments,
their squared-distance difference equals
\[
d_W^2(F,H_{L_n})-d_W^2(F,H_L)
=-2\langle F^{-1}-\mu_F,h_{L_n}-h_L\rangle.
\]
Cauchy--Schwarz and \eqref{eq:quantileerror} prove
\eqref{eq:budgeterror}.

To prove the risk bound, write $\Delta h=h_{L_n}-h_L$. Since both quantiles are
symmetric about $\mu$, $\Delta h(1-u)=-\Delta h(u)$ almost everywhere. The quantile
representation of the distortion riskmetric and the change of
variables $u\mapsto1-u$ give
\begin{equation*}
\rho_g(H_{L_n})-\rho_g(H_L)
=\int_0^1 g'(1-u)\Delta h(u)\,du=\frac12\int_0^1
\bigl[g'(1-u)-g'(u)\bigr]\Delta h(u)\,du=\int_0^1 \bar g'(1-u)\Delta h(u)\,du.
\end{equation*}
All integrals are finite by the Cauchy--Schwarz inequality.
Applying this inequality again and using \eqref{eq:quantileerror}
proves \eqref{eq:riskerror}.

Finally, antisymmetrization is an orthogonal projection in $L^2(0,1)$,
so $\delta_n\le\|{L_n-L}\|$. If $L_n\to L$ in $L^2(0,1)$, then $\delta_n\to0$, and
\eqref{eq:berror} implies $b_{L_n}\to b_L>0$. The standardized quantiles are
therefore well defined for all sufficiently large $n$, and
\eqref{eq:projectionerror}--\eqref{eq:riskerror} give the stated convergence.
\end{proof}

\end{document}